\documentclass[12pt]{article}

\usepackage[dvipsnames]{xcolor}
\usepackage{graphicx}
\usepackage{amsmath}        
\usepackage{amssymb}        
\usepackage{slashed}
\usepackage{bm}
\usepackage{here}
\usepackage{cite}

\def\mydate{April 18, 2025}
\def\ignore#1{{}}

\def\go{\rightarrow}
\def\dd{\partial}

\def\ep{{\epsilon}}

\def\SM{{\rm SM}}
\def\KK{{\rm KK}}
\def\EM{{\rm EM}}

\def\CKM{{\rm CKM}}

\def\onehalf{\hbox{$\frac{1}{2}$}}
\def\onethird{\hbox{$\frac{1}{3}$}}
\def\twothird{\hbox{$\frac{2}{3}$}}

\def\la{\langle}
\def\ra{\rangle}

\def\tree{{\rm tree}}

\def\mybig{\displaystyle \strut }

\def\myfrac#1#2{\frac{\mybig #1}{\mybig #2}}

\def\mymat#1#2{\begin{matrix}#1 \cr \noalign{\kern -2pt} #2\end{matrix}}

\def\mynoalign{\noalign{\kern 4pt}}
\def\mysnoalign{\noalign{\kern 3pt}}
\def\mytinynoalign{\noalign{\kern 2pt}}
\def\ignore#1{{}}

\makeatletter
\@addtoreset{equation}{section}
\makeatother

\begin{document}

\thispagestyle{empty}

{\small \noindent \mydate \hfill YITP-25-46}

\vskip 2.5cm

\baselineskip=30pt plus 1pt minus 1pt

\begin{center}
{\bf \Large Origin of CKM matrix and natural FCNC suppression 
in gauge-Higgs unification}\\ 
\end{center}


\baselineskip=22pt plus 1pt minus 1pt

\vskip 1.5cm

\begin{center}
\renewcommand{\thefootnote}{\fnsymbol{footnote}}
{\bf  Yutaka Hosotani$^{1}$\footnote[1]{hosotani@rcnp.osaka-u.ac.jp},
Shuichiro Funatsu$^2$, Hisaki Hatanaka$^3$}

{\bf  Yuta Orikasa$^4$ and Naoki Yamatsu$^5$}



\baselineskip=18pt plus 1pt minus 1pt

\vskip 10pt
{\small \it $^1$Research Center for Nuclear Physics, Osaka University,  Ibaraki, Osaka 567-0047, Japan}\\
{\small \it $^2$Ushiku, Ibaraki 300-1234, Japan} \\
{\small \it $^3$Osaka, Osaka 536-0014, Japan} \\
{\small \it $^4$Institute of Experimental and Applied Physics, Czech Technical University in Prague,} \\
{\small \it Husova 240/5, 110 00 Prague 1, Czech Republic} \\
{\small \it $^5$Yukawa Institute for Theoretical Physics, Kyoto University, Kyoto, Kyoto 606-8502, Japan} \\

\end{center}

\vskip 2.cm
\baselineskip=18pt plus 1pt minus 1pt

\begin{abstract}
In  the $SO(5) \times U(1) \times SU(3)$ gauge-Higgs unification in the Randall-Sundrum  warped  space
the mixing in the $W$ couplings in the quark sector is induced by masses of $SO(5)$ singlet fermions 
which are responsible for splitting masses of down-type quarks from those of up-type quarks in each generation.
We show that the observed Cabibbo-Kobayashi-Maskawa (CKM) matrix in the $W$ couplings is
reproduced within experimental errors.   Further flavor-changing neutral currents (FCNCs) in the quark sector
are shown to be naturally suppressed.
\end{abstract}

\newpage

\baselineskip=20pt plus 1pt minus 1pt
\parskip=0pt

\section{Introduction} 

Although  the standard model (SM) of  $SU(3)_C \times SU(2)_L \times U(1)_Y$ gauge theory has been 
successful in describing  phenomena at low energies, it has a gauge hierarchy problem 
when embedded in a larger theory such as grand unification.
As one way for solving  this problem  the gauge-Higgs unification  (GHU) scenario has been proposed in which 
gauge symmetry is dynamically broken by an Aharonov-Bohm (AB)  phase, $\theta_H$, in the fifth dimension.
The 4D Higgs boson is nothing but a 4D fluctuation mode of  $\theta_H$.  
A finite mass of the Higgs boson is dynamically generated by quantum effects of the AB phase $\theta_H$.
\cite{Hosotani1983, Davies1988, Hosotani1989, Davies1989,  HetrickHo1989, McLachlan1990, Hatanaka1998, 
Hatanaka1999,  Antoniadis2001, Takenaga2002, Kubo2002,   BurdmanNomura2003,  Csaki2003,  Scrucca2003, 
ACP2005, Cacciapaglia2006, Medina2007, HOOS2008, Serone2010, FHHOS2013, Yoon2018b,
GUTinspired2019a, FCNC2020a, GUTinspired2020b}

Among various GHU models the 
$SO(5)\times U(1) \times SU(3)$ GHU in the Randall-Sundrum (RS) 
warped  space\cite{RS1},   inspired by the $SO(11)$ gauge-Higgs grand unification model \cite{SO11GHGU},
has been extensively investigated.\cite{GUTinspired2019a, FCNC2020a, GUTinspired2020b}   
It has been shown that the grand unified theory (GUT) inspired GHU yields
nearly the same phenomenology at low energies as the SM.
It also gives many predictions to be confirmed by future experiments.
GHU models in the RS warped space predict large parity violation in the couplings of quarks and leptons
to Kaluza-Klein (KK) excited modes of gauge bosons, which can be  confirmed at electron-positron ($e^- e^+$) 
colliders.\cite{Funatsu2017a, Yoon2018a, Funatsu2019a, GUTinspired2020c, Irles2024}
Deviation from the SM can be seen in the processes of $W^- W^+$ production and single Higgs production as well.
\cite{Funatsu2023a, Yamatsu2023}
The GUT inspired GHU predicts a larger  $W$ boson mass than the SM,\cite{Wmass2023}
 just between the  SM value and the value reported by the CDF collaboration in 2022.\cite{CDF2022, ATLAS2023, CMS2024}
Generalization of the GHU scenario to grand unification has been attempted, too.
\cite{SO11GHGU, LimMaru2007, Kojima2011, HosotaniYamatsu2018,  Englert2020, Angelescu2022, Maru2022, Angelescu2023, MaruNago2024}
It is also recognized that GHU models on an orbifold such as the RS space exhibit the phenomenon of anomaly flow,
that is, the magnitude of gauge anomalies varies as the AB phase $\theta_H$ changes.  
Further there emerges a holographic formula for anomaly coefficients,  expressed in terms of the values of
the wave functions of gauge fields on the UV and IR branes.\cite{AnomalyFlow1, AnomalyFlow2}
The current status of GHU has been summarized in ref.\ \cite{YHbook}.

One of the issues remaining to be clarified in GHU is the flavor problem, namely the origin
of the Cabibbo-Kobayashi-Maskawa (CKM) matrix in the quark sector.\cite{FCNC2020a, Cacciapaglia2008, Adachi2010}
The current paper is devoted to solving this flavor problem in the GUT inspired GHU.
In the SM quark masses are generated by Yukawa couplings of quarks to the Higgs boson.
With general Yukawa couplings mass eigenstates are rotated from gauge eigenstates, which leads to
the nontrivial CKM matrix.  In GHU the $SU(2)_L \times U(1)_Y$ gauge symmetry is dynamically
broken to $U(1)_{EM}$ by the Hosotani mechanism so that couplings of quarks to the 4D Higgs boson
are controlled by the gauge principle.  One needs to explain how quark mass eigenstates deviate
from gauge eigenstates in GHU.
Previously an attempt   has been made in ref.\ \cite{FCNC2020a} to derive the CKM matrix
by introducing flavor mixing on the UV brane in the GUT inspired GHU.
Although the mixing in the $W$ boson couplings is qualitatively explained by brane interactions,
a realistic CKM matrix could not be obtained there.  In this paper we show that the observed 
CKM matrix is reproduced within experimental errors by incorporating general masses of $SO(5)$
singlet fermions which are responsible for splitting masses of down-type quarks from those of 
up-type quarks in the GUT inspired GHU.  It is also seen that flavor-changing neutral currents (FCNCs) 
are naturally suppressed.

In Section 2 the GUT inspired $SO(5) \times U(1)_X \times SU(3)_C$ GHU model with general
flavor mixing in the quark sector is described.  In Section 3 perturbation theory is developed by treating flavor
mixing as a small perturbation.  General formulas for wave functions are obtained.  
In Section 4 $W$ boson couplings of quarks are evaluated to show that the observed CKM matrix is reproduced.
In Section 5 $Z$ boson couplings are evaluated. It is shown that FCNCs are naturally suppressed.
Section 6 is devoted to a brief summary.  Basis functions for wave functions in the RS space are given in Appendix A.
$SO(5)$ generators in the spinor representation are given in Appendix B.
In Appendix C numerical values of coefficients in wave functions of $d$, $s$ and $b$ quarks are given.
In Appendix D $W$ and $Z$ couplings before flavor mixing are given.

\section{GUT inspired GHU with flavor mixing} 

The GUT inspired $SO(5) \times U(1)_X \times SU(3)_C (\equiv {\cal G})$ GHU was introduced in ref.\ \cite{GUTinspired2019a}.
It is defined in the RS warped space with its metric given by\cite{RS1}
\begin{align}
ds^2= G_{MN} \,  dx^M dx^N =
e^{-2\sigma(y)} \eta_{\mu\nu}dx^\mu dx^\nu+dy^2,
\label{RSmetric1}
\end{align}
where $M,N=0,1,2,3,5$, $\mu,\nu=0,1,2,3$, $y=x^5$, $\eta_{\mu\nu}=\mbox{diag}(-1,+1,+1,+1)$,
$\sigma(y)=\sigma(y+ 2L)=\sigma(-y)$, and $\sigma(y)=ky$ for $0 \le y \le L$.
In terms of the conformal coordinate $z=e^{ky}$ ($0 \le y \le L$, $1\leq z\leq z_L=e^{kL}$)
the metric is expressed as 
\begin{align}
ds^2=  \frac{1}{z^2} \bigg(\eta_{\mu\nu}dx^{\mu} dx^{\nu} + \frac{dz^2}{k^2}\bigg)~ .
\label{RSmetric-2}
\end{align}
The bulk region $0<y<L$ is anti-de Sitter (AdS) spacetime 
with a cosmological constant $\Lambda=-6k^2$, which is sandwiched by the
UV brane at $y=0$  and the IR brane at $y=L$.  
The warp factor $z_L$ is large, typically about $10^{11}$.
The KK mass scale is given by $m_{\rm KK}=\pi k/(z_L-1) \simeq \pi kz_L^{-1}$.

Gauge fields 
$A_M^{SO(5)}$,  $A_M^{U(1)_X}$ and $A_M^{SU(3)_C}$ of $SO(5) \times U(1)_X \times SU(3)_C$
with gauge couplings  $g_A$,  $g_B$ and $g_S$
satisfy the orbifold boundary conditions (BCs)
\begin{align}
&\begin{pmatrix} A_\mu \cr  A_{y} \end{pmatrix} (x,y_j-y) =
P_{j} \begin{pmatrix} A_\mu \cr  - A_{y} \end{pmatrix} (x,y_j+y)P_{j}^{-1}
\quad (j=0,1)
\label{BC-gauge1}
\end{align}
where $(y_0, y_1) = (0, L)$.  
Here $P_0=P_1 = P_{\bf 5}^{SO(5)} =\mbox{diag} (I_{4},-I_{1} )$ for $A_M^{SO(5)}$ in the vector representation and 
$P_0=P_1= 1$ for $A_M^{U(1)_X}$ and $A_M^{SU(3)_C}$.
The orbifold BCs break $SO(5)$ to $SO(4) \simeq SU(2)_L \times SU(2)_R$.
In the following we write $A_M^{SO(5)} = A_M$ and  $A_M^{U(1)_X} = B_M$ when no confusion arises.

The 4D Higgs boson doublet $\phi_H(x)$ is the zero mode contained in the 
$A_z = (kz)^{-1} A_y$ component;
\begin{align}
A_z^{(j5)} (x, z) &= \frac{1}{\sqrt{k}} \, \phi_j (x) u_H (z) + \cdots , ~~
u_H (z) = \sqrt{ \frac{2}{z_L^2 -1} } \, z ~, \cr
\noalign{\kern 5pt}
\phi_H(x) &= \frac{1}{\sqrt{2}} \begin{pmatrix} \phi_2 + i \phi_1 \cr \phi_4 - i\phi_3 \end{pmatrix} .
\label{4dHiggs}
\end{align}
Without loss of generality we assume $\la \phi_1 \ra , \la \phi_2 \ra , \la \phi_3 \ra  =0$ and  
$\la \phi_4 \ra \not= 0$, 
which is related to the Aharonov-Bohm (AB) phase $\theta_H$ in the fifth dimension by
$\la \phi_4 \ra  = \theta_H f_H$ where
\begin{align}
&f_H  = \frac{2}{g_w} \sqrt{ \frac{k}{L(z_L^2 -1)}} ~~,~ g_w = \frac{g_A}{\sqrt{L}} ~.
\label{fH1}
\end{align}

The matter content relevant in discussing the CKM matrix  is tabulated in Table \ref{Tab:matter}.
Fields $\Psi_{({\bf 3,4})}^{\alpha}$ ($\alpha=1,2,3$) are in the spinor representation of $SO(5)$, and
contain all quarks as zero modes.  In addition there are $SO(5)$ singlet fields $\Psi_{({\bf 3,1})}^{\pm \alpha}$
which have electromagnetic charge $ - \frac{1}{3} e$.  These fields satisfy the orbifold BCs 
\begin{align}
&\Psi_{({\bf 3,4})}^{\alpha} (x, y_j - y) = 
- P_{\bf 4}^{SO(5)} \gamma^5 \Psi_{({\bf 3,4})}^{\alpha} (x, y_j + y) ~, \cr
&\Psi_{({\bf 3,1})}^{\pm \alpha}  (x, y_j - y) =
\mp \gamma^5 \Psi_{({\bf 3,1})}^{\pm \alpha}  (x, y_j + y) ~.
\label{quarkBC1}
\end{align}
The bulk part of the action for the quark multiplets are given by
\begin{align}
S_{\rm bulk}^{\rm quark} &=  \int d^5x\sqrt{-\det G} \,
 \bigg\{ \sum_{\alpha = 1}^3 \Big( \overline{\Psi}{}_{\bf (3,4)}^\alpha   {\cal D} (c_{q^\alpha}) \Psi_{\bf (3,4)}^\alpha
+ \overline{\Psi}{}_{\bf (3,1)}^{+ \alpha}   {\cal D} (c_{D^\alpha} ) \Psi_{\bf (3,1)}^{+ \alpha} \cr
\noalign{\kern 5pt}
&\hskip -0.5cm
+ \overline{\Psi}{}_{\bf (3,1)}^{- \alpha}   {\cal D} (c_{D^\alpha} ) \Psi_{\bf (3,1)}^{- \alpha} \Big)
-  \sum_{\alpha, \beta=1}^3  \Big( m_{\alpha \beta}  \overline{\Psi}{}_{\bf (3,1)}^{+ \alpha} \Psi_{\bf (3,1)}^{- \beta} 
+ m_{\alpha \beta}^*  \overline{\Psi}{}_{\bf (3,1)}^{- \beta} \Psi_{\bf (3,1)}^{+ \alpha} \Big) \bigg\} ,
\label{fermionAction1}
\end{align} 
where $\overline{\Psi} = i \Psi^\dagger \gamma^0$ and the covariant derivative ${\cal D} (c)$ is given by 
\begin{align}
&{\cal D}(c)= \gamma^A {e_A}^M
\bigg( D_M+\frac{1}{8}\omega_{MBC}[\gamma^B,\gamma^C]  \bigg) -c \, \sigma'(y) ~, \cr
\noalign{\kern 5pt}
&D_M =  \dd_M - ig_S A_M^{SU(3)} -i g_A A_M^{SO(5)}  -i g_B Q_X A_M ^{U(1)} ~. 
\label{covariantD}
\end{align}

There is a brane scalar field $\hat \Phi_{({\bf 1}, {\bf 4})} (x)$ on the UV brane at $y=0$, which spontaneously
breaks $SU(2)_R \times U(1)_X$ to $U(1)_Y$ with $\la \hat \Phi_{({\bf 1}, {\bf 4})} \ra \not= 0$.
It has brane interactions with quark and lepton multiplets, which lead to splitting masses of down-type quarks
from those of up-type quarks, and also induce the inverse sea-saw mechanism for neutrinos.

\begin{table}[tbh]
\renewcommand{\arraystretch}{1.2}
\begin{center}
\caption{Parity assignment $(P_0, P_1)$ of quark multiplets in the bulk.
In the second column $\big( SU(3)_C, SO(5) \big)_{U(1)_X}$ content is shown.
In the third column $G_{22}=SU(2)_L\times SU(2)_R$ content is shown.
Brane scalar field $\Phi_{({\bf 1}, {\bf 4})}$ defined on the UV brane (at $y=0$) is   listed at the bottom row.
Note that $\la \hat \Phi_{[{\bf 1}, {\bf 2}]} \ra \not= 0$ while $\la \hat \Phi_{[{\bf 2}, {\bf 1}]} \ra = 0$.
}
\vskip 10pt
\begin{tabular}{cccccc}
\hline \hline 
Field & ${\cal G}$ & $G_{22}$ &Left-handed &Right-handed &Name\\
\hline
$\Psi_{({\bf 3,4})}^{\alpha}$ &$({\bf 3,4})_{\frac{1}{6}}$ &$\, [{\bf 2} , {\bf  1}] \,$
&$(+,+)$ &$(-,-)$ &$\begin{matrix} u~ & c & ~t \cr d~ & s & ~b\end{matrix}$\\
&&$[{\bf 1} , {\bf  2}]$ 
&$(-,-)$ &$(+,+)$ &$\begin{matrix} u'  & c' & t' \cr d' & s' & b' \end{matrix}$\\
$\Psi_{({\bf 3,1})}^{\pm \alpha}$ &$({\bf 3,1})_{-\frac{1}{3}}$ 
&$[{\bf 1} , {\bf  1}]$
&$(\pm ,\pm )$ &$(\mp , \mp )$ &$D^{\pm}_d ~ D^{\pm}_s ~ D^{\pm}_b$\\
$\hat \Phi_{({\bf 1}, {\bf 4})}$ &$({\bf 1,4})_{\frac{1}{2}}$ 
&$[{\bf 2} , {\bf  1}]$
&$\cdots$ & $\cdots$ &$\hat \Phi_{[{\bf 2}, {\bf 1}]}$\\
&&$[{\bf 1} , {\bf 2}]$ & $\cdots$ & $\cdots$ &$\hat \Phi_{[{\bf 1}, {\bf 2}]}$\\
\hline \hline
\end{tabular}
\label{Tab:matter}
\end{center}
\end{table}

The brane interactions of $\hat \Phi$ with quark multiplets are given by 
\begin{align}
&
 S_{\rm brane}^{\rm int}=
- \int d^5x\sqrt{-\det G} \, \delta(y) \, 
\bigg\{ \sum_{\alpha, \beta} \kappa_{\alpha\beta} \,
\overline{\Psi}{}_{({\bf 3,4})}^{\alpha}  \hat \Phi_{({\bf 1,4})}
 \Psi_{({\bf 3,1})}^{+\beta}  + {\rm H.c.} \bigg\}  ~.
\label{BraneInt1}
\end{align}
Nonvanishing $\la \hat \Phi_{[{\bf 1}, {\bf 2}]} \ra$ of $ \hat \Phi_{({\bf 1}, {\bf 4})}$ in Eq.\ (\ref{BraneInt1}) 
generates mass terms
\begin{align}
&S_{\rm brane\ mass}^{\rm fermion}=
\int d^5x\sqrt{-\det G} \, \delta(y)
 \Big\{  \sum_{\alpha, \beta} 2\mu_{\alpha\beta} \, 
\overline{d \,}{}_{R}^{\prime\alpha} D_{L}^{+\beta}  +\mbox{H.c.} \Big\} ~,
\label{braneFmass1}
\end{align}
where $(d^{\prime 1}, d^{\prime 2}, d^{\prime 3}) = (d', s', b')$ and 
$(D^{+ 1}, D^{+ 2}, D^{+ 3}) = (D_d^+, D_s^+, D_b^+)$. 

There are two sources for flavor mixing in the down-type quark sector, the $m_{\alpha\beta}$ terms in
Eq.\ (\ref{fermionAction1}) and $\mu_{\alpha\beta}$ terms in Eq.\ (\ref{braneFmass1}). 
In the previous paper (ref.\ \cite{FCNC2020a})  general form of brane interactions $\mu_{\alpha\beta}$,
but with $m_{\alpha\beta} = \delta_{\alpha\beta} m_{\alpha}$ was considered.  
Although the mixing in the $W$ boson couplings is induced by $\mu_{\alpha\beta}$ ($\alpha \not= \beta$),
it was not possible to explain the observed CKM matrix.
In the present paper we consider general mass terms for $D^{\pm \alpha}$ with 
$m_{\alpha\beta} \not= 0$ ($\alpha \not= \beta$).  We restrict ourselves to diagonal brane interactions
$\mu_{\alpha\beta} =  \delta_{\alpha\beta} \mu_\alpha$.

Manipulations are simplified in the twisted gauge \cite{Falkowski2007, HS2007}
defined by  an $SO(5)$ large gauge transformation
\begin{align}
&\tilde A_M (x,z) = \Omega A_M \Omega^{-1} 
- \frac{i}{g_A} \, \Omega \,\dd_M \Omega^{-1} ~, \cr
\noalign{\kern 5pt}
& \Omega (z)  = \exp \Big\{ i \theta (z) T^{45} \Big\}  ~,~~
\theta (z) = \theta_H \, \frac{z_L^2 - z^2}{z_L^2 - 1} ~, 
\label{twisted1}
\end{align}
where $T^{jk}$'s are $SO(5)$ generators and  
$A_M = 2^{-1/2} \sum_{1 \le j<k \le 5} A_M^{(jk)} T^{jk}$.
In the twisted gauge the background field vanishes ($\tilde \theta_H = 0$).
Boundary conditions at the UV brane are modified,  whereas boundary conditions at the IR brane remain 
the same as in the original gauge.  Quantities in the twisted gauge are denoted by the tilde sign $\tilde{~}$.

Up, charm, and top quarks are zero modes contained solely in $\Psi_{({\bf 3,4})}^{\alpha}$.
There arises no mixing in generation.
The mass spectrum $m_{q^{(n)}} = k \lambda_{q^{(n)}}$  ($q= u, c, t$) is determined by
\begin{align}
&S_L (1;  \lambda_{q^{(n)}}, c_q)  S_R (1;  \lambda_{q^{(n)}}, c_q) +\sin^2\onehalf \theta_H =0 ~,
\label{Up-quark-mass1}
\end{align}
where  basis functions $S_{L/R} (z, \lambda, c)$ and $C_{L/R} (z, \lambda, c)$ are given by (\ref{functionA2}).
(We follow the notation summarized in ref.\ \cite{YHbook}.)
The lowest modes $u^{(0)}$, $c^{(0)}$ and $t^{(0)}$ are $u$, $c$, and $t$ quarks.
With $m_u$, $m_c$ and $m_t$ given, the corresponding bulk mass parameters $c_u$, $c_c$ and $c_t$ are fixed,
and subsequently the mass spectra of their KK towers are determined.

For 5D fermion fields we define $\check \Psi = z^{-2} \Psi$.
The KK expansion of 5D $u (x,z)$ and $u' (x,z)$ fields in the twisted gauge  is given by
\begin{align}
& \begin{pmatrix} \tilde{\check u} \cr \tilde {\check u}' \end{pmatrix} = 
\sqrt{k} \sum_{n=0}^\infty \bigg\{ u^{(n)}_L (x) \begin{pmatrix} f^{u^{(n)}}_L (z) \cr g^{u^{(n)}}_L (z) \end{pmatrix}
+ u^{(n)}_R (x) \begin{pmatrix} f^{u^{(n)}}_R (z) \cr g^{u^{(n)}}_R (z) \end{pmatrix} \bigg\} , \cr
\noalign{\kern 5pt}
&\quad
\begin{pmatrix} f^{u^{(n)}}_L (z) \cr g^{u^{(n)}}_L (z) \end{pmatrix}=  \frac{1}{\sqrt{r_{u^{(n)} L}}}
\begin{pmatrix}\bar c_H C_L (z, \lambda_{u^{(n)}}, c_u) \cr
\noalign{\kern 5pt}
- i\bar s_H  \check S_L (z, \lambda_{u^{(n)}}, c_u) \end{pmatrix}, \cr
\noalign{\kern 5pt}
&\quad
\begin{pmatrix} f^{u^{(n)}}_R (z) \cr g^{u^{(n)}}_R (z) \end{pmatrix}=  \frac{1}{\sqrt{r_{u^{(n)} R}}}
\begin{pmatrix} \bar c_H S_R (z, \lambda_{u^{(n)}}, c_u) \cr
\noalign{\kern 5pt}
- i\bar s_H  \check C_R (z, \lambda_{u^{(n)}}, c_u) \end{pmatrix} ,
\label{wave-up1}
\end{align}
where $\check S_L$ and $\check C_R$ are defined in (\ref{functionA2}) and 
$(\bar c_H, \bar s_H) = (\cos \onehalf \theta_H, \sin \onehalf \theta_H)$. 
The normalization factor for each mode is determined by the condition
\begin{align}
\int_1^{z_L} dz \, \Big\{ |f (z) |^2 + |g (z) |^2 \Big\} = 1 \quad {\rm for~} 
\begin{pmatrix} f (z) \cr g (z) \end{pmatrix} .
\label{normalizationF1}
\end{align}
One can show that $r_{u^{(n)} L} = r_{u^{(n)} R} $.  Similar formulas hold for charm and top towers.

For down-type quarks $d^\alpha$, $d^{\prime \alpha}$, $D^{+ \alpha}$ and $D^{- \alpha}$ intertwine with each other.
Equations of motion in the original gauge are given by
\begin{align}
&\sigma^\mu \dd_\mu \begin{pmatrix} \check d^\alpha_L \cr \check d^{\prime \alpha}_L \end{pmatrix}
- k \Big( D_- (c_{q^\alpha} )  + \frac{i}{2} \theta' (z) \tau^1 \Big) 
 \begin{pmatrix} \check d^\alpha_R \cr \check d^{\prime \alpha}_R \end{pmatrix} = 0 ~,  \cr
\noalign{\kern 5pt}
&\bar \sigma^\mu \dd_\mu \begin{pmatrix} \check d^\alpha_R \cr \check d^{\prime \alpha}_R \end{pmatrix}
- k \Big( D_+ (c_{q^\alpha} )  -  \frac{i}{2} \theta' (z) \tau^1 \Big) 
 \begin{pmatrix} \check d^\alpha_L \cr \check d^{\prime \alpha}_L \end{pmatrix} 
 = 2 \mu_\alpha \delta (y)  \begin{pmatrix} 0 \cr \check D^{+ \alpha}_L \end{pmatrix} ,  \cr
\noalign{\kern 5pt}
&\sigma^\mu \dd_\mu \begin{pmatrix} \check D^{+ \alpha}_L \cr \check D^{- \alpha}_L \end{pmatrix}
- k D_- (c_{D^\alpha} ) \begin{pmatrix} \check D^{+ \alpha}_R \cr \check D^{- \alpha}_R \end{pmatrix}
- \frac{1}{z} \begin{pmatrix} 0 & m_{\alpha\beta} \cr m_{\beta\alpha}^* & 0 \end{pmatrix}
 \begin{pmatrix} \check D^{+ \beta}_R \cr \check D^{- \beta}_R \end{pmatrix}
 = 2 \mu_\alpha^* \delta (y) \begin{pmatrix} \check d^{\prime \alpha}_R \cr 0\end{pmatrix} , \cr
\noalign{\kern 5pt}
&\bar \sigma^\mu \dd_\mu \begin{pmatrix} \check D^{+ \alpha}_R \cr \check D^{- \alpha}_R \end{pmatrix}
- k D_+ (c_{D^\alpha} ) \begin{pmatrix} \check D^{+ \alpha}_L\cr \check D^{- \alpha}_L \end{pmatrix}
- \frac{1}{z} \begin{pmatrix} 0 & m_{\alpha\beta} \cr m_{\beta\alpha}^* & 0 \end{pmatrix}
 \begin{pmatrix} \check D^{+ \beta}_L \cr \check D^{- \beta}_L \end{pmatrix} = 0 ~.
 \label{downtypeEq1}
\end{align}
Here $D_\pm (c) = \pm (d/dz) + (c/z)$.
Brane interactions in (\ref{braneFmass1}) give rise to $\delta (y)$ terms on the right-hand side of the equations above.

$d^\alpha_R$, $d^{\prime \alpha}_L$, $D^{+ \alpha}_R$ and $D^{- \alpha}_L$ are parity odd at $y=0$.
Because of the brane interactions $d^{\prime \alpha}_L$ and $D^{+ \alpha}_R$ become discontinuous at $y=0$ as
\begin{align}
&\check d^{\prime \alpha}_L \big|_{y=\ep} = - \check d^{\prime \alpha}_L \big|_{y=-\ep} 
= - \mu_\alpha \check  D^{+ \alpha}_L \big|_{y=0} ~, \cr
\noalign{\kern 5pt}
&\check D^{+ \alpha}_R \big|_{y=\ep} = - \check D^{+ \alpha}_R \big|_{y=- \ep} 
= \mu_\alpha^* \check d^{\prime \alpha}_R \big|_{y=0} ~, \cr
\noalign{\kern 5pt}
&\check d^\alpha_R \big|_{y=0}  =  \check D^{- \alpha}_L \big|_{y=0} = 0 ~.
\label{downBC1}
\end{align}
This is seen by integrating the equations in the $y$ coordinate from $- \ep$ to $\ep$ and taking the limit $\ep \go 0$
with $k D_\pm (c) = e^{-\sigma(y)} [ \pm (\dd/\dd y) + c \sigma '(y) ]$ in the $y$-coordinate.

Each eigenmode of a mass $m=k\lambda$ has $x$-dependence $\check \Psi_{L/R} (x,z) = f_{L/R} (x) \check \psi_{L/R} (z)$ where
$\sigma^\mu \dd_\mu f_L(x) = k\lambda  f_R (x)$ and $\bar \sigma^\mu \dd_\mu f_R(x) = k\lambda  f_L (x)$.
Equations for mode functions ($\check \psi_{L/R} (z)$ part of each field) in the twisted gauge in the bulk region 
$1 < z< z_L$ ($0 < y < L$) become
\begin{align}
&\lambda \begin{pmatrix} \tilde{\check d}^\alpha_L \cr \tilde{\check d}^{\prime \alpha}_L \end{pmatrix}
- D_- (c_{q^\alpha} ) \begin{pmatrix} \tilde{\check d}^\alpha_R \cr \tilde{\check d}^{\prime \alpha}_R \end{pmatrix} = 0 ~,  \cr
\noalign{\kern 5pt}
&\lambda \begin{pmatrix} \tilde{\check d}^\alpha_R \cr \tilde{\check d}^{\prime \alpha}_R \end{pmatrix}
-D_+ (c_{q^\alpha} ) 
 \begin{pmatrix} \tilde{\check d}^\alpha_L \cr \tilde{\check d}^{\prime \alpha}_L \end{pmatrix} = 0 ~, \cr
\noalign{\kern 5pt}
&\lambda \begin{pmatrix} \check D^{+ \alpha}_L \cr \check D^{- \alpha}_L \end{pmatrix}
-  D_- (c_{D^\alpha} ) \begin{pmatrix} \check D^{+ \alpha}_R \cr \check D^{- \alpha}_R \end{pmatrix}
- \frac{1}{z} \begin{pmatrix} 0 & \tilde m_{\alpha\beta} \cr \tilde m_{\beta\alpha}^* & 0 \end{pmatrix}
\begin{pmatrix} \check D^{+ \beta}_R \cr \check D^{- \beta}_R \end{pmatrix} = 0 ~, \cr
\noalign{\kern 5pt}
&\lambda \begin{pmatrix} \check D^{+ \alpha}_R \cr \check D^{- \alpha}_R \end{pmatrix}
-  D_+ (c_{D^\alpha} ) \begin{pmatrix} \check D^{+ \alpha}_L\cr \check D^{- \alpha}_L \end{pmatrix}
- \frac{1}{z} \begin{pmatrix} 0 & \tilde m_{\alpha\beta} \cr\tilde  m_{\beta\alpha}^* & 0 \end{pmatrix}
 \begin{pmatrix} \check D^{+ \beta}_L \cr \check D^{- \beta}_L \end{pmatrix} = 0 ~,
 \label{downtypeEq2}
\end{align}
where $\tilde m_{\alpha\beta} = m_{\alpha\beta}/k$.  Boundary conditions are 
\begin{align}
\underline{{\rm at~} z= 1 } :~~~
&\bar c_H \tilde{\check d}^\alpha_R + i \bar s_H \tilde{\check d}^{\prime \alpha}_R = 0 ~, \cr
& i \bar s_H \tilde{\check d}^\alpha_L + \bar c_H  \tilde{\check d}^{\prime \alpha}_L 
+ \mu_\alpha  \check  D^{+ \alpha}_L = 0 ~,\cr
& \check  D^{+ \alpha}_R - \mu_\alpha^* (  i \bar s_H \tilde{\check d}^\alpha_R 
+  \bar c_H  \tilde{\check d}^{\prime \alpha}_R ) = 0 ~,  \cr
&\check  D^{- \alpha}_L = 0 ~, \cr
\underline{{\rm at~} z= z_L } :~
& \tilde{\check d}^\alpha_R =  \tilde{\check d}^{\prime \alpha}_L =  \check  D^{+ \alpha}_R = \check  D^{- \alpha}_L = 0 ~.
\label{downBC2}
\end{align}
Note that the other half of boundary conditions are obtained from Eq.\  (\ref{downtypeEq2}) with (\ref{downBC2}).
For instance, at $z=z_L$,  $\tilde{\check d}^\alpha_R =0$ implies $D_+ (c_{q^\alpha} )  \, \tilde{\check d}^\alpha_L =0$.

\section{Perturbation theory}

It is not easy to rigorously solve Eq.\ (\ref{downtypeEq2}) with general  $\tilde m_{\alpha\beta}$.
The case of diagonal masses $\tilde m_{\alpha\beta} = m_\alpha \delta_{\alpha\beta}$ has been solved
in ref.\ \cite{GUTinspired2019a}.
In this paper we treat cases where off-diagonal elements of $\tilde m_{\alpha\beta}$ ($\alpha \not= \beta$) 
are much smaller than diagonal elements $\tilde m_{\alpha\alpha} \equiv \tilde m_\alpha$.

To develop a perturbation theory we rewrite the equations in Eq.\ (\ref{downtypeEq2}) in the form familiar
in quantum mechanics.  To simplify expressions we write
\begin{align}
&d^\alpha = \begin{pmatrix} \vec d^\alpha_{L}  \cr \vec d^\alpha_{R}  \end{pmatrix}, ~~ 
\vec d^\alpha_{L/R} = \begin{pmatrix}  \tilde{\check d}^\alpha_{L/R} \cr  \tilde{\check d}^{\prime \alpha}_{L/R} \end{pmatrix} ,   \cr
\noalign{\kern 5pt}
&D^\alpha = \begin{pmatrix} \vec D^\alpha_{L} \cr  \vec D^\alpha_{R} \end{pmatrix} , ~~
\vec D^\alpha_{L/R} = \begin{pmatrix}  \check D^{+ \alpha}_{L/R} \cr \check D^{- \alpha}_{L/R} \end{pmatrix} ,
\label{dDvector1}
\end{align}
and introduce
\begin{align}
K^q_\alpha \, d^\alpha &= \begin{pmatrix} 0 & D_- (c_{q^\alpha}) \cr D_+ (c_{q^\alpha}) & 0 \end{pmatrix}
 \begin{pmatrix} \vec d^\alpha_{L}  \cr \vec d^\alpha_{R}  \end{pmatrix} , \cr
 \noalign{\kern 5pt}
K^D_\alpha \, D^\alpha &=  \begin{pmatrix} 0 & D_- (c_{D^\alpha}) + z^{-1} \hat m_\alpha \cr 
D_+ (c_{D^\alpha})  +  z^{-1} \hat m_\alpha & 0 \end{pmatrix}
 \begin{pmatrix} \vec D^\alpha_{L}  \cr \vec D^\alpha_{R}  \end{pmatrix} , \cr
\noalign{\kern 5pt}
\hat m_\alpha &= \begin{pmatrix} 0 &  \tilde m_\alpha \cr \tilde m_\alpha^* & 0 \end{pmatrix} .
\label{defK}
\end{align}
Each eigenmode of Eq.\ (\ref{downtypeEq2}) has 24 components. 
Eq.\ (\ref{downtypeEq2}) is written as 
\begin{align}
&(K + V) \, \Psi = \lambda \, \Psi ~,
\label{downtypeEq3}
\end{align}
where
\begin{align}
&K = \begin{pmatrix} K^q_1 \cr & K^q_2 \cr && K^q_3 \cr &&& K^D_1 \cr &&&& K^D_2 \cr &&&&& K^D_3 \end{pmatrix} , \quad
 \Psi = \begin{pmatrix} d^1 \cr d^2 \cr d^3 \cr D^1 \cr D^2 \cr D^3 \end{pmatrix} , \cr
\noalign{\kern 5pt}
&V =  \begin{pmatrix}0 \cr &0 \cr && 0 \cr 
&&& 0 & V_{12} & V_{13} \cr &&& V_{21} & 0 & V_{23} \cr &&& V_{31} & V_{32} & 0 \end{pmatrix} , ~~
 V_{\alpha\beta}  = \frac{1}{z} \begin{pmatrix} &&& \tilde m_{\alpha\beta} \cr && \tilde m_{\beta\alpha}^* \cr
&  \tilde m_{\alpha\beta} \cr   \tilde m_{\beta\alpha}^*  \end{pmatrix} .
 \label{downtypeEq4}
\end{align}
We treat $V$ as a small perturbation and determine $\Psi$ and $\lambda$ in perturbation theory.
Note that the boundary conditions in Eq.\ (\ref{downBC2}) do not depend on mass perturbation parameters $\tilde m_{\alpha\beta}$
($\alpha \not= \beta$). 

In the absence of $V$, or to the zeroth order, the equations have been completely solved.  In each generation
$K^q_\alpha d^\alpha = \lambda \, d^\alpha $ and $K^D_\alpha D^\alpha = \lambda \, D^\alpha$.
We write wave functions of the $n$-th eigenmode with $\lambda_{\alpha^{(n)}}$ as 
\begin{align}
&d^\alpha = \begin{pmatrix} \vec d^\alpha_{L}  \cr \vec d^\alpha_{R}  \end{pmatrix}
= \begin{pmatrix} f^{\alpha^{(n)}}_L \cr  g^{\alpha^{(n)} }_L \cr  f^{\alpha^{(n)}}_R \cr  g^{\alpha^{(n)} }_R \end{pmatrix} , ~~
 D^\alpha =  \begin{pmatrix} \vec D^\alpha_{L} \cr  \vec D^\alpha_{R} \end{pmatrix} 
 =\begin{pmatrix}  h^{\alpha^{(n)}}_L \cr  k^{\alpha^{(n)} }_L \cr  h^{\alpha^{(n)}}_R \cr  k^{\alpha^{(n)} }_R \end{pmatrix} .
\label{downWave1}
\end{align}
They are given by
\begin{align}
\begin{pmatrix} f^{\alpha^{(n)}}_L (z) \cr \mynoalign g^{\alpha^{(n)}}_L (z) \cr \mynoalign 
h^{\alpha^{(n)}}_L (z) \cr \mynoalign k^{\alpha^{(n)}}_L (z)\end{pmatrix} 
&= \frac{1}{ \sqrt{r_{\alpha^{(n)}}} }
\begin{pmatrix}  A_{\alpha^{(n)}} C_L(z)\cr
\mynoalign
B_{\alpha^{(n)}}S_L(z)\cr
\mynoalign
a_{\alpha^{(n)}}{\cal C}_{L2}(z)
+b_{\alpha^{(n)}}{\cal C}_{L1}(z) \cr
\mynoalign
a_{\alpha^{(n)}}{\cal S}_{L1}(z)
+b_{\alpha^{(n)}}{\cal S}_{L2}(z) \cr  \end{pmatrix} \equiv  | \psi^{\alpha^{(n)}}_L \ra ~, \cr
\noalign{\kern 5pt}
\begin{pmatrix} f^{\alpha^{(n)}}_R (z) \cr \mynoalign g^{\alpha^{(n)}}_R (z) \cr \mynoalign 
h^{\alpha^{(n)}}_R (z) \cr \mynoalign k^{\alpha^{(n)}}_R (z)\end{pmatrix} 
&= \frac{1}{ \sqrt{r_{\alpha^{(n)}}} }
\begin{pmatrix}  A_{\alpha^{(n)}} S_R(z)\cr
\mynoalign
B_{\alpha^{(n)}}C_R(z)\cr
\mynoalign
a_{\alpha^{(n)}}{\cal S}_{R2}(z)
+b_{\alpha^{(n)}}{\cal S}_{R1}(z) \cr
\mynoalign
a_{\alpha^{(n)}}{\cal C}_{R1}(z)
+b_{\alpha^{(n)}}{\cal C}_{R2}(z) \cr  \end{pmatrix} \equiv  | \psi^{\alpha^{(n)}}_R \ra ~,   \cr
\noalign{\kern 5pt}
C_{L/R}(z) &=C_{L/R} (z; \lambda_{\alpha^{(n)}} , c_{q^\alpha})  , \cr
S_{L/R}(z) &=S_{L/R} (z; \lambda_{\alpha^{(n)}} , c_{q^\alpha}) , \cr
{\cal C}_{L/R j}(z) &= {\cal C}_{L/R j}(z;\lambda_{\alpha^{(n)}}, c_{D^\alpha}, \tilde m_\alpha) , ~\cr
{\cal S}_{L/R j}(z) &= {\cal S}_{L/R j}(z;\lambda_{\alpha^{(n)}}, c_{D^\alpha}, \tilde m_\alpha) , 
\label{downWave2}
\end{align}
where
\begin{align}
B_{\alpha^{(n)}} &= 
 i \, \frac{\bar c_H S_R(1) }{\bar s_H C_R(1)} \, 
A_{\alpha^{(n)}} ~, \cr
 \noalign{\kern 5pt}
a_{\alpha^{(n)}} &= - i \,
\frac{\mu_d S_R(1) \, {\cal S}_{L2}(1 )}{\bar s_H [ {\cal S}_{L1}  {\cal S}_{R1} (1) - {\cal S}_{L2}  {\cal S}_{R2} (1)] } \, 
A_{\alpha^{(n)}} ~, \cr
\noalign{\kern 5pt}
b_{\alpha^{(n)}} &= - 
\frac{ {\cal S}_{L1}(1 )} {{\cal S}_{L2}(1)} \,
a_{\alpha^{(n)}} ~.
\label{downWave3}
\end{align}
Without loss of generality we have taken $\tilde m_\alpha^* = \tilde m_\alpha$.  
${\cal C}_{L/R j}$ and ${\cal S}_{L/R j}$ are defined in Eq.\ (\ref{MassiveFermion1}).
The boundary conditions  in Eq.\ (\ref{downBC2}) are satisfied with 
\begin{align}
&\Big( S_L S_R +\sin^2\frac{\theta_H}{2} \Big)
 \big({\cal S}_{L1} {\cal S}_{R1}  -{\cal S}_{L2} {\cal S}_{R2} \big)   \cr
 \noalign{\kern 5pt}
 &\hskip 1.cm
+|\mu_\alpha |^2 C_R S_R
 \big( {\cal S}_{L1} {\cal C}_{L1}  -{\cal S}_{L2}  {\cal C}_{L2}  \big) \Big|_{z=1} =0 ~.
\label{downSpectrum1}
\end{align}
The value of $c_{D^\alpha}$ is determined by Eq.\  (\ref{downSpectrum1}) with a given down-type quark mass 
($m_{d^\alpha} = k \lambda_{\alpha^{(0)}}$).
Brane interactions ($\mu_\alpha \not= 0$) are important in splitting  $m_d, m_s, m_b$ from $m_u, m_c, m_t$, respectively.
Further it results in almost entirely suppressing 
$W$ couplings of right-handed components of quarks.
Once $c_{D^\alpha}$ is fixed, the spectrum $\{ \lambda_{\alpha^{(n)}} \}$ of each KK tower is determined by 
Eq.\ (\ref{downSpectrum1}).
Wave functions are normalized by
\begin{align}
&\int_1^{z_L} dz \big\{ | f_L |^2 + | g_L |^2 + | h_L |^2 + | k_L |^2 \big\} \cr
&= \int_1^{z_L} dz \big\{ | f_R |^2 + | g_R|^2 + | h_R |^2 + | k_R |^2 \big\} = 1
\label{normalizationDown1}
\end{align}
for each mode $\alpha^{(n)}$.  Note that the KK tower of $\{ \alpha^{(n)} \}$, say,  for $\alpha=1 = d$, contains
both $d^{(n)} (n \ge 0)$ and $D_d^{(n)} (n \ge 1)$ modes.

Let us denote a state vector for the  $\alpha^{(n)}$ mode as
\begin{align}
&K \, | \psi_{\alpha^{(n)}} \ra = \lambda_{\alpha^{(n)}} \,  | \psi_{\alpha^{(n)}} \ra ~, \cr
\noalign{\kern 5pt}
&| \psi_{\alpha^{(n)}} \ra = \begin{pmatrix}  | \psi^{\alpha^{(n)}}_L \ra \cr  | \psi^{\alpha^{(n)}}_R \ra \end{pmatrix} , \cr
\noalign{\kern 5pt}
&\la  \psi_{\alpha^{(n)}}  |  \psi_{\beta^{(m)}} \ra =  
\la  \psi^{\alpha^{(n)}}_L  |  \psi^{\beta^{(m)}}_L \ra + \la  \psi^{\alpha^{(n)}}_R  |  \psi^{\beta^{(m)}}_R \ra 
= 2 \, \delta_{\alpha\beta} \, \delta_{nm} ~. 
\label{downState1}
\end{align}
It is understood that wave functions of $\beta\not= \alpha$ components of the state vector $| \psi_{\alpha^{(n)}} \ra$ vanish.
Then matrix elements of $V$ are given by
\begin{align}
\la  \psi_{\alpha^{(n)}}  | V |  \psi_{\beta^{(\ell)}} \ra & = 
\begin{cases} 0 &{\rm for~} \alpha = \beta  \cr
\tilde m_{\alpha\beta} \, W_{\alpha^{(n)}, \beta^{(\ell)}} + \tilde m_{\beta\alpha}^*  \, (W_{\beta^{(\ell)}, \alpha^{(n)}} )^* 
 &{\rm for~} \alpha \not= \beta 
\end{cases} ~, \cr
\noalign{\kern 10pt}
W_{\alpha^{(n)}, \beta^{(\ell)}}  &= \int_1^{z_L} \frac{dz}{z} \Big( h_L^{\alpha^{(n)} *} \, k_R^{\beta^{(\ell)}} 
+ h_R^{\alpha^{(n)} *} \, k_L^{\beta^{(\ell)}}  \Big) ~.
\label{Velement1}
\end{align}

We determine eigen functions $\Psi$ in Eq.\  (\ref{downtypeEq3}) for $d, s, b$ quarks in perturbation theory.
We write
\begin{align}
| \Psi^d \ra  & = | \psi_{1^{(0)}} \ra  +\sum_{\alpha=1}^3 \sum_n  c^d_{\alpha ^{(n)}} | \psi_{\alpha ^{(n)}} \ra  ~, \cr
\noalign{\kern 5pt}
| \Psi^s\ra & = | \psi_{2^{(0)}} \ra  +\sum_{\alpha=1}^3 \sum_n  c^s_{\alpha ^{(n)}} | \psi_{\alpha ^{(n)}} \ra  ~, \cr
\noalign{\kern 5pt}
| \Psi^b\ra & = | \psi_{3^{(0)}} \ra  +\sum_{\alpha=1}^3 \sum_n  c^b_{\alpha ^{(n)}} | \psi_{\alpha ^{(n)}} \ra  ~.
\label{downWave4}
\end{align}
It turns out that contributions of KK modes, $\alpha^{(n)} (n \ge 1)$, to $W$ and $Z$ couplings of quarks 
are small. 
As is seen in the next section, the observed $W$ couplings indicate 
$\tilde m_\alpha \gg |\tilde m_{12}| \gg  |\tilde m_{23}| \gg  |\tilde m_{13}|$, and one can suppose that
\begin{align}
&|\tilde m_{12}|, |\tilde m_{21}| = O(\ep) ~, \cr
&|\tilde m_{23}|, |\tilde m_{32}| = O(\ep^2) ~, \cr
&|\tilde m_{13}|, |\tilde m_{31}| = O(\ep^3) ~, 
\label{Vorder1}
\end{align}
where $\ep$ is a small  parameter.
First we note that first-order corrections to eigenvalues $\lambda_d, \lambda_s$ and $\lambda_b$ vanish.
\begin{align}
&\lambda_d = \lambda_d^{(0)} + \lambda_d^{(1)} + \cdots , ~~
 \lambda_d^{(0)} = \lambda_{1^{(0)}}, ~~  \lambda_d^{(1)} = 0 ~, \cr
&\lambda_s = \lambda_s^{(0)} + \lambda_s^{(1)} + \cdots , ~~
 \lambda_s^{(0)} = \lambda_{2^{(0)}}, ~~  \lambda_s^{(1)} = 0 ~, \cr
&\lambda_b = \lambda_b^{(0)} + \lambda_b^{(1)} + \cdots , ~~
 \lambda_b^{(0)} = \lambda_{3^{(0)}}, ~~  \lambda_b^{(1)} = 0 ~.
  \label{downMass2}
\end{align}

Let us expand the coefficient as $ c^d_{\alpha ^{(n)}} = c^{d \, (1)}_{\alpha ^{(n)}} + c^{d\, (2)}_{\alpha ^{(n)}} + \cdots$ etc.,
where $c^{d \, (j)}_{\alpha ^{(n)}}$ is the $j$-th order correction in perturbation theory.
Relevant coefficients are given by 
\begin{align}
&c^{d \, (1)}_{1 ^{(n)}} = 0 ~, ~~ 
c^{d \, (2)}_{1 ^{(0)}} = - \frac{1}{2} \sum_n | c^{d \, (1)}_{2 ^{(n)}} |^2 - \frac{1}{2} \sum_n | c^{d \, (1)}_{3 ^{(n)}} |^2 
= O(\ep^2) , \cr
\noalign{\kern 5pt}
&c^{d \, (1)}_{2 ^{(n)}} = \frac{\la  \psi_{2^{(n)}}  | V |  \psi_{1^{(0)}} \ra}{2 (\lambda_{1^{(0)}} - \lambda_{2^{(n)}} )} = O(\ep) ,~
c^{d \, (2)}_{2 ^{(0)}} \sim  \frac{\la  \psi_{2^{(0)}}  | V |  \psi_{3^{(0)}} \ra \la  \psi_{3^{(0)}}  | V |  \psi_{1^{(0)}} \ra}
{4 (\lambda_{1^{(0)}} - \lambda_{2^{(0)}} ) (\lambda_{1^{(0)}} - \lambda_{3^{(0)}} )} 
= O(\ep^5) , \cr
\noalign{\kern 5pt}
&c^{d \, (1)}_{3 ^{(n)}} = \frac{\la  \psi_{3^{(n)}}  | V |  \psi_{1^{(0)}} \ra}{2 (\lambda_{1^{(0)}} - \lambda_{3^{(n)}} )} = O(\ep^3) , ~
c^{d \, (2)}_{3 ^{(0)}} \sim  \frac{\la  \psi_{3^{(0)}}  | V |  \psi_{2^{(0)}} \ra \la  \psi_{2^{(0)}}  | V |  \psi_{1^{(0)}} \ra}
{4 (\lambda_{1^{(0)}} - \lambda_{3^{(0)}} ) (\lambda_{1^{(0)}} - \lambda_{2^{(0)}} )} = O(\ep^3) , \cr
\noalign{\kern 5pt}
&c^{s \, (1)}_{2 ^{(n)}} = 0 ~, ~~ 
c^{s \, (2)}_{2 ^{(0)}} =  - \frac{1}{2} \sum_n | c^{s \, (1)}_{1 ^{(n)}} |^2    - \frac{1}{2} \sum_n | c^{s \, (1)}_{3 ^{(n)}} |^2 
= O(\ep^2) , \cr
\noalign{\kern 5pt}
&c^{s \, (1)}_{1 ^{(n)}} = \frac{\la  \psi_{1^{(n)}}  | V |  \psi_{2^{(0)}} \ra}{2 (\lambda_{2^{(0)}} - \lambda_{1^{(n)}} )} = O(\ep) ,~
c^{s \, (2)}_{1 ^{(0)}} \sim  \frac{\la  \psi_{1^{(0)}}  | V |  \psi_{3^{(0)}} \ra \la  \psi_{3^{(0)}}  | V |  \psi_{2^{(0)}} \ra}
{4 (\lambda_{2^{(0)}} - \lambda_{1^{(0)}} ) (\lambda_{2^{(0)}} - \lambda_{3^{(0)}} )} 
= O(\ep^5) , \cr
\noalign{\kern 5pt}
&c^{s \, (1)}_{3 ^{(n)}} = \frac{\la  \psi_{3^{(n)}}  | V |  \psi_{2^{(0)}} \ra}{2 (\lambda_{2^{(0)}} - \lambda_{3^{(n)}} )} = O(\ep^2) , ~
c^{s \, (2)}_{3 ^{(0)}} \sim  \frac{\la  \psi_{3^{(0)}}  | V |  \psi_{1^{(0)}} \ra \la  \psi_{1^{(0)}}  | V |  \psi_{2^{(0)}} \ra}
{4 (\lambda_{2^{(0)}} - \lambda_{3^{(0)}} ) (\lambda_{2^{(0)}} - \lambda_{1^{(0)}} )}
= O(\ep^4), \cr
\noalign{\kern 5pt}
&c^{b \, (1)}_{3 ^{(n)}} = 0~, ~~
c^{b \, (2)}_{3 ^{(0)}} =  - \frac{1}{2} \sum_n | c^{b \, (1)}_{1 ^{(n)}} |^2    - \frac{1}{2} \sum_n | c^{b \, (1)}_{2 ^{(n)}} |^2 
= O(\ep^4) , \cr
\noalign{\kern 5pt}
&c^{b \, (1)}_{1 ^{(n)}} = \frac{\la  \psi_{1^{(n)}}  | V |  \psi_{3^{(0)}} \ra}{2 (\lambda_{3^{(0)}} - \lambda_{1^{(n)}} )} = O(\ep^3) , ~
c^{b \, (2)}_{1 ^{(0)}} \sim  \frac{\la  \psi_{1^{(0)}}  | V |  \psi_{2^{(0)}} \ra \la  \psi_{2^{(0)}}  | V |  \psi_{3^{(0)}} \ra}
{4 (\lambda_{3^{(0)}} - \lambda_{1^{(0)}} ) (\lambda_{3^{(0)}} - \lambda_{2^{(0)}} )} = O(\ep^3) , \cr
\noalign{\kern 5pt}
&c^{b \, (1)}_{2 ^{(n)}} = \frac{\la  \psi_{2^{(n)}}  | V |  \psi_{3^{(0)}} \ra}{2 (\lambda_{3^{(0)}} - \lambda_{2^{(n)}} )} = O(\ep^2) , ~
c^{b \, (2)}_{2 ^{(0)}} \sim  \frac{\la  \psi_{2^{(0)}}  | V |  \psi_{1^{(0)}} \ra \la  \psi_{1^{(0)}}  | V |  \psi_{3^{(0)}} \ra}
{4 (\lambda_{3^{(0)}} - \lambda_{2^{(0)}} ) (\lambda_{3^{(0)}}- \lambda_{1^{(0)}} )}
= O(\ep^4). 
\label{downWave5}
\end{align}
A factor 2 in the denominator of, say, $c^{d \, (1)}_{2 ^{(n)}}$ appears due to the normalization taken in Eq.\  (\ref{downState1}).
Since $m_d, m_s, m_b \ll m_\KK$, contributions coming from KK excited ($\alpha^{(n)}, n \ge 1)$ modes 
to the coefficients $c^{d,s,b}_{\alpha^{(n)}}$ are suppressed.  
Furthermore it is seen that 
\begin{align}
& c^{d \, (1)}_{2 ^{(0)}}  = - \big(c^{s \, (1)}_{1 ^{(0)}} \big)^* ~, \cr
\noalign{\kern 5pt}
& c^{s \, (1)}_{3 ^{(0)}}  = - \big( c^{b \, (1)}_{2 ^{(0)}} \big)^* ~,  \cr
\noalign{\kern 5pt}
& c^{d \, (1)}_{3 ^{(0)}}  = - \big(c^{b \, (1)}_{1 ^{(0)}} \big)^* ~, 
\label{downWave6}
\end{align}
and
\begin{align}
& c^{s \, (2)}_{1 ^{(0)}}  + \big(c^{d \, (2)}_{2 ^{(0)}} \big)^* + \big(c^{d \, (1)}_{3 ^{(0)}} \big)^* c^{s \, (1)}_{3 ^{(0)}} \sim 0 ~, \cr
\noalign{\kern 5pt}
& c^{b \, (2)}_{2 ^{(0)}}  + \big(c^{s \, (2)}_{3 ^{(0)}} \big)^* + \big(c^{s \, (1)}_{1 ^{(0)}} \big)^* c^{b \, (1)}_{1 ^{(0)}}  \sim  0 ~, \cr
\noalign{\kern 5pt}
& c^{b \, (2)}_{1 ^{(0)}}  + \big(c^{d \, (2)}_{3 ^{(0)}} \big)^* + \big(c^{d \, (1)}_{2 ^{(0)}} \big)^* c^{b \, (1)}_{2 ^{(0)}} \sim  0 ~. 
\label{downWave7}
\end{align}
These properties are important in discussing $W$ and $Z$ couplings of quarks.

\section{$W$ boson couplings} 

Gauge couplings of fermions are contained in the covariant derivative $D_M$ in Eq.\ (\ref{covariantD}).
Let $ \{ T^{a_L}, T^{a_R}, T^{\hat b} ; a=1 \sim 3, b= 1 \sim 4 \}$ 
be generators of $SO(5)$.
The $W$ boson field, $W_\mu (x)$,  is contained in 
$\sum_{a=1}^2 ( A_\mu^{a_L}  T^{a_L} + A_\mu^{a_R}  T^{a_R} + A_\mu^{\hat a} T^{\hat a} )$.
Wave functions of the $W$-boson tower in the twisted gauge are given by \cite{YHbook}
\begin{align}
&\frac{1}{\sqrt{2 k}} \begin{pmatrix} \tilde A_\mu^{1_L} + i \tilde A_\mu^{2_L} \cr
\tilde A_\mu^{1_R} + i \tilde A_\mu^{2_R} \cr 
\tilde A_\mu^{\hat 1} + i \tilde A_\mu^{\hat 2} \end{pmatrix} 
=  \sum_{n=0}^\infty W_\mu^{ (n)} (x) \begin{pmatrix}  h^L_{W^{(n)}} (z) \cr \mysnoalign
 h^R_{W^{(n)}} (z) \cr  \mysnoalign  \hat h_{W^{(n)}} (z) \end{pmatrix} + \cdots , \cr
\noalign{\kern 5pt}
& \begin{pmatrix}  h^L_{W^{(n)}} (z) \cr \mysnoalign
 h^R_{W^{(n)}} (z) \cr  \mysnoalign  \hat h_{W^{(n)}} (z) \end{pmatrix} =
 \frac{1}{\sqrt{2 \, r_{W^{(n)}}}}
\begin{pmatrix} (1 + c_H) \, C(z; \lambda_{W^{(n)}}) \cr 
(1 - c_H) \, C(z; \lambda_{W^{(n)}})  \cr  
\sqrt{2} \, s_H \check S(z; \lambda_{W^{(n)}}) \end{pmatrix}  , \cr
\noalign{\kern 5pt}
&\quad c_H = \cos \theta_H ~, ~~ s_H = \sin \theta_H ~, 
\label{WbosonWave1}
\end{align}
where the spectrum $m_{W^{(n)}} = k \lambda_{W^{(n)}}$ is determined by
 $2 S C' (1; \lambda_{W^{(n)}}) +s_H^2  \lambda_{W^{(n)}}  =0$.
 $C(z; \lambda)$ and $S(z; \lambda)$ are defined in Eq.\  (\ref{functionA1}), and 
 $\check S(z; \lambda) = [C(1;\lambda)/S(1;\lambda)] S(z; \lambda)$.
Wave functions are normalized by
\begin{align}
&\int_1^{z_L}  \frac{dz}{z} \Big\{ ( |h^L_{W^{(n)}} |^2 + |h^R_{W^{(n)}} |^2 + | \hat h_{W^{(n)}} |^2 \Big\} = 1 ~.
\label{Wnormalization}
\end{align}
The $W$ boson field is $W_\mu (x) = W_\mu^{(0)}  (x)$.

The $W$ interaction in the twisted gauge is evaluated from
\begin{align}
{\cal L}^W_{\rm int} &= - ig_A \sum_\alpha  \int_1^{z_L} \frac{dz}{\sqrt{k}} \, 
 \overline{\tilde{\check \Psi}}{}_{\bf (3,4)}^\alpha 
\gamma^\mu
\sum_{a=1}^2 ( \tilde A_\mu^{a_L}  T^{a_L} + \tilde A_\mu^{a_R}  T^{a_R} + \tilde  A_\mu^{\hat a} T^{\hat a} )
\tilde{\check \Psi}{}_{\bf (3,4)}^\alpha  
\label{Wfermion1}
\end{align}
by inserting the wave functions of quarks and $W$ boson presented above.
For  quarks one finds that
\begin{align}
{\cal L}^W_{\rm int} &= - i  \frac{g_w}{\sqrt{2}} W_\mu^\dagger  \bigg\{
( \bar u_L, \bar c_L, \bar t_L) \gamma^\mu   \hat g^{W q}_L \begin{pmatrix} d_L \cr s_L \cr b_L \end{pmatrix}
+ (L \go R) \bigg\}  \cr
\noalign{\kern 5pt}
\hat g^{W q}_L &= 
\begin{pmatrix}  \hat g^{Wud}_{L} & \hat g^{Wus}_{L}  &\hat g^{Wub}_{ L} \cr
\hat g^{Wcd}_{L} & \hat g^{Wcs}_{L}  &\hat g^{Wcb}_{L} \cr
\hat g^{Wtd}_{L} & \hat g^{Wts}_{L}  &\hat g^{Wtb}_{L} \end{pmatrix} ,
\label{Wcoupling1}
\end{align}
where 
\begin{align}
&( \hat g^{Wud}_{L} ,   \hat g^{Wus}_{L} ,  \hat g^{Wub}_{ L}  ) = 
\sum_n ( \delta_{n\, 0} + c^d_{d^{(n)}} ,  c^s_{d^{(n)}}  , c^b_{d^{(n)}} ) \,\hat g^W_{L \, u^{(0)} d^{(n)}} ~, \cr
&( \hat g^{Wcd}_{L} ,   \hat g^{Wcs}_{L} ,  \hat g^{Wcb}_{ L}  ) = 
\sum_n(  c^d_{s^{(n)}} ,   \delta_{n\, 0} + c^s_{s^{(n)}}  , c^b_{s^{(n)}} ) \, \hat g^W_{L \, c^{(0)} s^{(n)}} ~, \cr
&( \hat g^{Wtd}_{L} ,   \hat g^{Wts}_{L} ,  \hat g^{Wtb}_{ L}  ) = 
\sum_n(  c^d_{b^{(n)}} ,  c^s_{b^{(n)}}  ,  \delta_{n\, 0}+c^b_{b^{(n)}} ) \, g^W_{L \, t^{(0)} b^{(n)}}  ~, \cr
\noalign{\kern 5pt}
&\hat g^W_{L \, u^{(0)} d^{(n)}} = G_W[ (h^L, h^R, \hat h)_{W^{(0)}}; ( f,g)^{u^{(0)}}_{L} ,   (f, g)^{d^{(n)}}_{L} ] ~, \cr
&\hat g^W_{L \, c^{(0)} s^{(n)}} = G_W[ (h^L, h^R, \hat h)_{W^{(0)}}; ( f,g)^{c^{(0)}}_{L} ,   (f, g)^{s^{(n)}}_{L} ] ~, \cr
&\hat g^W_{L \, t^{(0)} b^{(n)}} = G_W[ (h^L, h^R, \hat h)_{W^{(0)}}; ( f, g)^{t^{(0)}}_{L} ,   (f, g)^{b^{(n)}}_{L} ] ~, \cr
\noalign{\kern 8pt}
&G_W [(h^L, h^R, \hat h)_\alpha; (f, g)_1, (f, g)_2]  \cr
\noalign{\kern 2pt}
&\quad
= \sqrt{kL} \int_1^{z_L} dz \, \Big\{ h^{L*}_\alpha   f_1^*   f_2 +  h^{R*}_\alpha  g_1^*   g_2
+  \frac{i}{\sqrt{2}} \,  \hat h^*_\alpha  ( f_1^*  g_2 -  g_1^*  f_2 ) \Big\} ~.
\label{Wcoupling2}
\end{align}
Note that $\{ 1^{(n)} \} =  \{ d^{(n)} \} +  \{ D_d^{(n)} \}$ etc.
Contributions from  $D_d^{(n)}$,  $D_s^{(n)}$ and  $D_b^{(n)}$ modes are small and omitted in the above.
Right-handed couplings are very small; $\hat g^{Wud}_R = O(10^{-12})$,  $\hat g^{Wcs}_R = O(10^{-9})$ and 
$\hat g^{Wtb}_R = O(10^{-5})$.  
The CKM matrix in GHU is given by
\begin{align}
V_\CKM = \begin{pmatrix} V_{ud} & V_{us} & V_{ub} \cr V_{cd} & V_{cs} & V_{cb} \cr V_{td} & V_{ts} & V_{tb} \end{pmatrix}
&= \frac{1}{\hat g_L^{W \nu_e e}} \, \hat g^{W q}_L
\label{CKMghu1}
\end{align}
where the couplings are normalized by the $\nu_e e$ coupling $ \hat g^{W\nu_e e}_{L}$.  

In the Wolfenstein parametrization of the CKM matrix $V_\CKM$ is expressed as 
\begin{align}
&V_\CKM^{\rm Wolfenstein} = \begin{pmatrix} 1 - \lambda^2/2 & \lambda & A \lambda^3 (\rho - i \eta) \cr
- \lambda & 1 - \lambda^2/2 & A \lambda^2 \cr
A \lambda^3 (1 - \rho - i \eta) & - A\lambda^2 & 1 \end{pmatrix} + O(\lambda^4) 
\label{Wolfenstein1}
\end{align}
where $\lambda = 0.225$, $A=0.826$, $\rho = 0.163$ and $\eta = 0.361$.\cite{PDG2024}
In the SM, $V_\CKM^\SM$ is unitary.
In GHU there are infinitely many KK excited states, and $V_\CKM$ given in Eq.\ (\ref{CKMghu1}) is
a matrix truncated to the zero modes of quark towers so that $V_\CKM$ is expected to be 
unitary only approximately.

In the following we present numerical values for  $\theta_H=0.1$ and $m_\KK = 13\,$TeV.
(Predicted values for the case with $\theta_H=0.08$ and $m_\KK = 16\,$TeV are listed 
in Table \ref{Tab:parameter} below.)
For these values $z_L = 3.832 \times 10^{11}$ and $k= 1.586 \times 10^{15} \,$GeV with $m_Z$ as an input.
The bulk mass parameters of the quark multiplets are determined, with $m_u$, $m_c$ and $m_t$ given, to be
$(c_u, c_c, c_t) = (- 0.8591, -0.7191, - 0.2745)$.  Brane interaction parameters and diagonal elements of 
mass parameters in the down-quark sector
are taken to be  $(\mu_1, \mu_2, \mu_3) = (0.1, 0.1, 1.0)$ and $\tilde m_1= \tilde m_2 =\tilde m_3 = 0.8$,  
with which bulk mass parameters of $D^{\pm \alpha}$ fields become 
$(c_{D_d}, c_{D_s}, c_{D_b}) = (0.4241, 0.4560, 0.6720)$ with $m_d$, $m_s$, and $m_b$ given.
$\sin^2 \theta_W^0 = g_B^2/(g_A^2 + 2 g_B^2)$ is determined such that the Fermi constant $G_F$ is consistently
reproduced.\cite{Wmass2023}.  $\sin^2 \theta_W^0 (m_Z) = 0.23035$ at the $m_Z$ scale.
The resulting $W$ couplings before the mixing are $ \hat g^{W\nu_e e}_{L} = 0.997647$, 
$\hat g_{Wud}^{(0)}  = 0.997645$, $\hat g_{Wcs}^{(0)}  = 0.997643$ and $\hat g_{Wtb}^{(0)}  = 0.997969$.

With the hierarchy observed in the Wolfenstein parametrization of $V_\CKM$ in Eq.\ (\ref{Wolfenstein1})
in mind, we expect the hierarchy in Eq.\ (\ref{Vorder1}) in $\tilde m_{\alpha\beta} \, (\alpha\not=\beta)$.
We find that parameter values 
\begin{align}
\tilde m_{12} &= \tilde m_{21}^* = 1.18 \times 10^{-2} ~, \cr
\tilde m_{23} &= \tilde m_{32}^* = 8.10 \times 10^{-5} ~, \cr
\tilde m_{13} &= \tilde m_{31}^* = (1.52 - 3.74 \, i ) \times 10^{-6} ~
\label{tilmass1}
\end{align}
yield, to the second order in mass perturbation, 
\begin{align}
V_\CKM &= 
\begin{pmatrix} 0.9758& 0.2163 + 0.0076 i & (1.532 - 3.372 \, i ) \times 10^{-3} \cr
-0.2163 + 0.0075 i  & 0.9750 & ( 4.076 - 0.001 i ) \times 10^{-2}\cr
(7.412 - 3.372 \,  i) \times 10^{-3} & ( - 4.106 - 0.075 i )\times 10^{-2}& 0.9995 \end{pmatrix},
\label{CKMghu2}
\end{align}
which is close to the value obtained in the SM  global fit.
Details of the evaluation have been described in Appendices C and D.
The angles of the unitarity triangle are
\begin{align}
\alpha &= \arg \bigg( - \frac{V_{td} V_{tb}^*}{V_{ud} V_{ub}^*} \bigg) = 1.5703 = 89.97 {\,}^\circ , \cr
\noalign{\kern 5pt}
\beta &= \arg \bigg( - \frac{V_{cd} V_{cb}^*}{V_{td} V_{tb}^*} \bigg) = 0.3927 = 22.50  {\,}^\circ , \cr
\noalign{\kern 5pt}
&\qquad \sin 2 \beta = 0.7071 ~, \cr 
\noalign{\kern 5pt}
\gamma &= \arg \bigg( - \frac{V_{ud} V_{ub}^*}{V_{cd} V_{cb}^*} \bigg) = 1.1786 = 67.53{\,}^\circ .
\label{unitarity1}
\end{align}
Experimental values are $\alpha_{\rm exp} = (84.1^{+4.5}_{-3.8} )^\circ $, $\sin 2\beta_{\rm exp} = 0.709 \pm 0.011$
and $\gamma_{\rm exp} = (65.7 \pm 3.0 )^\circ$.\cite{PDG2024}
In the SM global fit $\alpha_\SM= 91.6^\circ $, $\sin 2\beta_\SM  = 0.713$ and $\gamma_\SM = 65.7^\circ$.  
One sees that the values in Eq.\ (\ref{tilmass1}) reproduce $V_\CKM$ consistent with the observation.
We do not need fine tuning of the parameters $\tilde m_{ab}$.  When one takes $\tilde m_{12} = 1.19 \times 10^{-2}$
instead of $1.18 \times 10^{-2}$,
for instance, then $V_{us}$ and $\sin 2 \beta$ vary by about 1\%.

\section{$Z$ boson couplings} 

The spectrum of the $Z$ boson tower, $\{ m_{Z^{(n)}} = k \lambda_{Z^{(n)}}  \}$,  is determined by 
\begin{align}
&2 S C' (1; \lambda_{Z^{(n)}}) +  \lambda_{Z^{(n)}} \frac{\sin^2 \theta_H }{\cos^2 \theta_W^0}=0 ~, \cr
\noalign{\kern 5pt}
&\sin \theta_W^0  = \frac{g_B}{\sqrt{g_A^2 + 2 g_B^2}} ~.
\label{Zspectrum1}
\end{align}
The $Z$ boson field is contained in 
$A_\mu^{3_L}$, $ A_\mu^{3_R} $, $A_\mu^{\hat 3} $, and $B_\mu $.
Wave functions of the $Z$-boson tower in the twisted gauge are given by \cite{YHbook}
\begin{align}
&\frac{1}{\sqrt{k}} \begin{pmatrix} \tilde A_\mu^{3_L}  \cr \mytinynoalign \tilde A_\mu^{3_R}\cr 
\mytinynoalign  \tilde A_\mu^{\hat 3}  \cr B_\mu  \end{pmatrix} 
= \sum_{n=0}^\infty Z_\mu^{ (n)} (x) \begin{pmatrix}  h^L_{Z^{(n)}} (z) \cr \mysnoalign
 h^R_{Z^{(n)}} (z) \cr  \mysnoalign    \hat h_{Z^{(n)}} (z) \cr  \mysnoalign h^B_{Z^{(n)}} (z) \end{pmatrix} + \cdots , \cr
\noalign{\kern 5pt}
& \begin{pmatrix}  h^L_{Z^{(n)}} (z) \cr \mysnoalign
 h^R_{Z^{(n)}} (z) \cr  \mysnoalign    \hat h_{Z^{(n)}} (z) \cr  \mysnoalign h^B_{Z^{(n)}} (z) \end{pmatrix}  = 
  \begin{pmatrix}  h^{L, su2}_{Z^{(n)}} (z) \cr \mysnoalign
 h^{R, su2}_{Z^{(n)}} (z) \cr  \mysnoalign    \hat h^{su2}_{Z^{(n)}} (z) \cr   0 \end{pmatrix}
 - \sin \theta_W^0  \begin{pmatrix} \sin \theta_W^0 \cr   \sin \theta_W^0 \cr     0 \cr 
\sqrt{1 - 2 \sin^2 \theta_W^0} \end{pmatrix} h^{em}_{Z^{(n)}} (z) , \cr
\noalign{\kern 5pt}
&\hskip 0.7cm
 \begin{pmatrix}  h^{L, su2}_{Z^{(n)}} (z) \cr \mysnoalign
 h^{R, su2}_{Z^{(n)}} (z) \cr  \mysnoalign    \hat h^{su2}_{Z^{(n)}} (z)  \end{pmatrix} =
\frac{1}{\sqrt{ \, 2 \, r_{Z^{(n)}} }} 
\begin{pmatrix} (1 + c_H) \,C(z, \lambda_{Z^{(n)}} ) \cr 
(1 -c_H) \,C(z, \lambda_{Z^{(n)}} ) \cr  
  \sqrt{2}s_H  \check S (z, \lambda_{Z^{(n)}}  ) \end{pmatrix} , \cr
\noalign{\kern 5pt}
&\hskip 1.cm
h^{em}_{Z^{(n)}} (z)  = \sqrt{\frac{2}{r_{Z^{(n)}}}} \, C(z, \lambda_{Z^{(n)}} )~.
\label{ZbosonWave1}
\end{align}
Wave functions are normalized by
\begin{align}
&\int_1^{z_L}  \frac{dz}{z} \Big\{ ( |h^L_{Z^{(n)}} |^2 + |h^R_{Z^{(n)}} |^2 + | \hat h_{Z^{(n)}} |^2  +  |h^B_{Z^{(n)}} |^2 \Big\} = 1 ~.
\label{Znormalization}
\end{align}
The $Z$ boson field is $Z_\mu (x) = Z_\mu^{(0)}  (x)$.

The $Z$ interaction in the twisted gauge is evaluated from
\begin{align}
{\cal L}_{\rm int}^{Z} &= - ig_A \sum_J  \int_1^{z_L} \frac{dz}{\sqrt{k}} \cr
\noalign{\kern 5pt}
&\quad \times 
 \overline{\tilde{\check \Psi}}{}^J  \gamma^\mu
\Big(   \tilde A_\mu^{3_L}  T^{3_L} + \tilde A_\mu^{3_R}  T^{3_R} + \tilde  A_\mu^{\hat 3} T^{\hat 3}
+ \frac{g_B}{g_A}  B_\mu Q_X \Big) \tilde{\check \Psi}^J  
\label{Zfermion1}
\end{align}
by inserting the wave functions of quarks and $Z$ boson presented above.
Here the sum $\sum_J$ extends over $\tilde{\check \Psi}{}_{\bf (3,4)}^\alpha$ and $\tilde{\check \Psi}{}_{\bf (3,1)}^{\pm \alpha}$.
The $Z$ couplings of quarks are written as 
\begin{align}
{\cal L}^Z_{\rm int} &= - i \frac{g_w}{\cos \theta_W^0} Z_\mu  \bigg\{
( \bar u_L, \bar c_L, \bar t_L) \gamma^\mu \, \hat g^{Z u}_{L}   \begin{pmatrix} u_L \cr c_L \cr t_L \end{pmatrix}
+ (L \go R)  \cr
\noalign{\kern 5pt}
&\hskip 2.5cm
+ ( \bar d_L, \bar s_L, \bar b_L) \gamma^\mu  \, \hat g^{Z d }_{L}  \begin{pmatrix} d_L \cr s_L \cr b_L \end{pmatrix} 
+ (L \go R) \bigg\} ~. \cr
\noalign{\kern 5pt}
\hat g^{Z u}_{L/R} &= \begin{pmatrix}  \hat g^{Zuu}_{L/R} & 0  &0 \cr  0 & \hat g^{Zcc}_{L/R}  &0 \cr
0&0 &\hat g^{Ztt}_{L/R} \end{pmatrix},  ~~
\hat g^{Z d}_{L/R} = \begin{pmatrix}  \hat g^{Zdd}_{L/R} & \hat g^{Zds}_{L/R}  &\hat g^{Zdb}_{ L/R} \cr
\hat g^{Zsd}_{L/R} & \hat g^{Zss}_{L/R}  &\hat g^{Zsb}_{L/R} \cr
\hat g^{Zbd}_{L/R} & \hat g^{Zbs}_{L/R}  &\hat g^{Zbb}_{L/R} \end{pmatrix} 
\label{Zcoupling1} 
\end{align}

There is no mixing in the up-type quark sector.  
Let $\hat g^Z_{L/R \, d^{(n)} d^{(m)}}$, $\hat g^Z_{L/R \, s^{(n)} s^{(m)}}$ and $\hat g^Z_{L/R \, b^{(n)} b^{(m)}}$
be the $Z$ couplings of
the $n$-th and $m$-th KK modes of $d$, $s$ and $b$ quarks before the mixing, respectively.  
Then one finds, in each generation,  that 
\begin{align}
\hat g^{Z uu}_{L/R} ~~~ &= \hat g^{Z uu, su2}_{L/R} - \sin^2 \theta_W^0 \, \hat g^{Z uu, em}_{L/R} ~, \cr
\noalign{\kern 5pt}
\hat g^{Z uu, su2}_{L/R} ~ &= \cos \theta_W^0 T^3_u \, 
G_W[ (h^L, h^R, \hat h)^{su2}_{Z^{(0)}};  (f, g)^{u^{(0)}}_{L/R} , (f, g)^{u^{(0)}}_{L/R} ] , \cr
\noalign{\kern 5pt}
\hat g^{Z uu, em}_{L/R}  ~&=  \cos \theta_W^0 Q_u \,
G_\gamma^u [ h^{em}_{Z^{(0)}} ;   (f, g)^{u^{(0)}}_{L/R},   (f, g)^{u^{(0)}}_{L/R} ] , 
\label{Zcoupling2a}
\end{align}
and 
\begin{align}
\hat g^Z_{L/R \, d^{(n)} d^{(m)}}&= \hat g^{Z , su2}_{L/R\, d^{(n)} d^{(m)}} - \sin^2 \theta_W^0 \, \hat g^{Z , em}_{L/R\, d^{(n)} d^{(m)}} ~, \cr
\noalign{\kern 5pt}
 \hat g^{Z , su2}_{L/R\, d^{(n)} d^{(m)}}  &= \cos \theta_W^0 T^3_d \, 
G_W[ (h^L, h^R, \hat h)^{su2}_{Z^{(0)}};  (f, g)^{d^{(n)}}_{L/R} , (f, g)^{d^{(m)}}_{L/R} ] , \cr
\noalign{\kern 5pt}
\hat g^{Z , em}_{L/R\, d^{(n)} d^{(m)}}  &=  \cos \theta_W^0 Q_d \,
G_\gamma^d [ h^{em}_{Z^{(0)}} ;   (f, g, h, k)^{d^{(n)}}_{L/R},   (f, g,h,k)^{d^{()}}_{L/R} ] , 
\label{Zcoupling2b}
\end{align}
where
\begin{align}
&G_\gamma^u [h_\gamma ; (f, g)_1 , (f, g)_2 ]  
= \sqrt{kL} \int_1^{z_L} dz  \, h_\gamma  \big( f_1^*  f_2 + g_1^* g_2  \big) , \cr
\noalign{\kern 5pt}
&G_\gamma^d [h_\gamma ; (f, g, h, k)_1 , (f, g, h, k)_2 ]  \cr
\noalign{\kern 5pt}
&\hskip 1.cm
= \sqrt{kL} \int_1^{z_L} dz  \, h_\gamma  \big( f_1^*  f_2 + g_1^* g_2  +  h_1^*  h_2 + k_1^* k_2 \big) , \cr
\noalign{\kern 5pt}
&(T_u^3, T_d^3) = (\onehalf, - \onehalf) ~, ~~ (Q_u, Q_d) = (\twothird, - \onethird) ~.
\label{Zcoupling3}
\end{align}
For $\theta_H=0.1$, $m_\KK = 13\,$TeV and  $\sin^2 \theta_W^0 = 0.23035$ we find that
\begin{align}
&\frac{1}{\hat g_L^{W \nu_e e}} \begin{pmatrix} \hat g^{Z uu}_L \cr \hat g^{Z cc}_L \cr  \hat g^{Z tt}_L \end{pmatrix}
= \begin{pmatrix} 0.34606 \cr 0.34606 \cr 0.34639 \end{pmatrix} , ~~
\frac{1}{\hat g_L^{W \nu_e e}} \begin{pmatrix} \hat g^{Z uu}_R \cr \hat g^{Z cc}_R\cr  \hat g^{Z tt}_R \end{pmatrix}
= \begin{pmatrix} - 0.15394 \cr  - 0.15394 \cr  - 0.15361 \end{pmatrix} ,  \cr
\noalign{\kern 5pt}
&\frac{1}{\hat g_L^{W \nu_e e}} 
\begin{pmatrix} \hat  g^Z_{L \, d^{(0)} d^{(0)}} \cr \hat  g^Z_{L \, s^{(0)} s^{(0)}} \cr  \hat  g^Z_{L \, b^{(0)} b^{(0)}} \end{pmatrix}
= \begin{pmatrix} - 0.42304 \cr - 0.42304  \cr - 0.42303  \end{pmatrix} , ~~
\frac{1}{\hat g_L^{W \nu_e e}} 
\begin{pmatrix} \hat  g^Z_{R \, d^{(0)} d^{(0)}} \cr \hat g^Z_{R \, s^{(0)} s^{(0)}} \cr  \hat  g^Z_{R \, b^{(0)} b^{(0)}} \end{pmatrix}
= \begin{pmatrix} 0.07697 \cr 0.07697   \cr 0.07698  \end{pmatrix} .
\label{Zcoupling4}
\end{align}
In the SM  with $\sin^2 \theta_W^\SM = 0.2312$,  
$\hat g_L^{W \nu_e e} = 1$, 
$\hat g^{Z}_{u\, L} = 0.3459$ and $\hat g^{Z}_{u\, R} = - 0.1541$.

The $Z$ couplings $\hat g^{Z\beta\gamma}_{L/R} = (\hat g^{Z d}_{L/R})^{\beta\gamma}$ 
($\beta, \gamma = d, s, b$) in Eq.\ (\ref{Zcoupling1}) are expressed, with (\ref{downWave4}) inserted,  as 
\begin{align}
&\hat g^{Z\beta\gamma}_{L/R} = \hat g^{Z\gamma\beta \, *}_{L/R}
= \sum_{\alpha} \sum_{n,m} (\delta_{\alpha \beta} \delta_{n0} + c^{\beta ~ *}_{\alpha^{(n)}} ) \, 
\hat g^Z_{L/R \, \alpha^{(n)}\alpha^{(m)}}  (\delta_{\alpha \gamma} \delta_{m 0} + c^\gamma_{\alpha^{(m)}} ) ~.
\label{ZcouplingA1}
\end{align}
We evaluate these couplings to the second order in perturbation theory.
The coefficients $c^\beta_{\alpha^{(n)}}$ appear in different patterns for diagonal and off-diagonal elements.
Let us first take $\hat g^{Z\, dd}_L$ as an example.  One finds that
\begin{align}
\hat g^{Z\, dd}_L &=  \sum_{n,m} (\delta_{n0} + c^{d ~ *}_{1^{(n)}} ) \,  \hat g^Z_{L \,1^{(n)}1^{(m)}}  ( \delta_{m 0} + c^d_{1^{(m)}} ) \cr
\noalign{\kern 5pt}
&\quad
+   \sum_{n,m} c^{d ~ *}_{2^{(n)}}  \,  \hat g^Z_{L \,2^{(n)}2^{(m)}}  c^d_{2^{(m)}} 
+   \sum_{n,m} c^{d ~ *}_{3^{(n)}}  \,  \hat g^Z_{L \,3^{(n)}3^{(m)}}  c^d_{3^{(m)}}  \cr
\noalign{\kern 5pt}
&= (1 +   c^{d\, (2)} _{1^{(0)}} + c^{d\, (2) *} _{1^{(0)}} ) \hat g^Z_{L \,d^{(0)}d^{(0)}} \cr
\noalign{\kern 5pt}
&\quad 
+  \sum_{n,m} c^{d \, (1) *}_{2^{(n)}}  \,  \hat g^Z_{L \,2^{(n)}2^{(m)}}  c^{d \, (1)}_{2^{(m)}} 
+   \sum_{n,m} c^{d \, (1)  *}_{3^{(n)}}  \,  \hat g^Z_{L \,3^{(n)}3^{(m)}}  c^{d \, (1)}_{3^{(m)}} + \cdots  \cr
\noalign{\kern 5pt}
&= \hat g^Z_{L \,d^{(0)}d^{(0)}}  +  \big| c^{d \, (1)}_{2^{(0)}} \big|^2 \delta \hat g^Z_{L \,s^{(0)}s^{(0)} }
+  \big| c^{d \, (1)}_{3^{(0)}} \big|^2 \delta \hat g^Z_{L \,b^{(0)}b^{(0)} }  + \cdots ,
\label{ZcouplingA2}
\end{align}
where $\delta \hat g^Z_{L \,s^{(0)}s^{(0)} }  = \hat g^Z_{L \,s^{(0)}s^{(0)}} -  \hat g^Z_{L \,d^{(0)}d^{(0)}}$ and 
 $\delta \hat g^Z_{L \,b^{(0)}b^{(0)} }  = \hat g^Z_{L \,b^{(0)}b^{(0)}} -  \hat g^Z_{L \,d^{(0)}d^{(0)}}$.
In the last equality we have made use of the relations in Eq.\ (\ref{downWave5}{) and have omitted negligible terms.
As $\hat g^Z_{L \,s^{(0)}s^{(0)}},  \hat g^Z_{L \,b^{(0)}b^{(0)}}  \sim \hat g^Z_{L \,d^{(0)}d^{(0)}}$, the correction is very small
and one finds $\hat g^{Z\, dd}_L \sim  \hat g^Z_{L \,d^{(0)}d^{(0)}}$.  Thorough evaluation gives 
\begin{align}
&\frac{1}{\hat g_L^{W \nu_e e}} 
\begin{pmatrix} \hat  g^{Z\, dd}_L \cr \hat   g^{Z\, ss}_L\cr  \hat   g^{Z\, bb}_L  \end{pmatrix}
= \begin{pmatrix} - 0.42293 \cr - 0.42291  \cr - 0.42303  \end{pmatrix} , ~~
\frac{1}{\hat g_L^{W \nu_e e}} 
\begin{pmatrix} \hat   g^{Z\, dd}_R \cr \hat g^{Z\, ss}_R \cr  \hat  g^{Z\, bb}_R \end{pmatrix}
= \begin{pmatrix} 0.07697 \cr 0.07697   \cr 0.07698  \end{pmatrix} .
\label{ZcouplingA3}
\end{align}
In the SM,  
$\hat g^{Z}_{d\, L} = -0.4229$ and $\hat g^{Z}_{d\, R} = 0.0771$.

For off-diagonal elements,  take $\hat g^{Z\, db}_L$ as an example. 
\begin{align}
\hat g^{Z\, db}_L &=  \sum_{n,m} (\delta_{n0} + c^{d ~ *}_{1^{(n)}} ) \,  \hat g^Z_{L \,1^{(n)}1^{(m)}} c^b_{1^{(m)}}  \cr
\noalign{\kern 5pt}
&\qquad
+   \sum_{n,m} c^{d ~ *}_{2^{(n)}}  \,  \hat g^Z_{L \,2^{(n)}2^{(m)}}  c^b_{2^{(m)}} 
+   \sum_{n,m} c^{d ~ *}_{3^{(n)}}  \,  \hat g^Z_{L \,3^{(n)}3^{(m)}}  (\delta_{m 0} +  c^b_{3^{(m)}} ) \cr
\noalign{\kern 5pt}
&=  \sum_{n} \Big\{ ( c^{b\, (1)}_{1^{(n)}} + c^{b\, (2)}_{1^{(n)}} ) \hat g^Z_{L \,1^{(0)}1^{(n)}} +
  ( c^{d\, (1)*}_{3^{(n)}} + c^{d\, (2)*}_{3^{(n)}} )\hat g^Z_{L \,3^{(n)}3^{(0)}} \Big\} \cr
\noalign{\kern 5pt}
&\hskip 2.cm
+  \sum_{n,m} c^{d \, (1) *}_{2^{(n)}}  \,  \hat g^Z_{L \,2^{(n)} 2^{(m)}}    c^{b \, (1)}_{2^{(m)}} \cr
\noalign{\kern 5pt}
&=  ( c^{b\, (1)}_{1^{(0)}} + c^{d\, (1)*}_{3^{(0)}} ) \,  \hat g^Z_{L \,d^{(0)}d^{(0)}}
+  ( c^{b\, (2)}_{1^{(0)}} + c^{d\, (2)*}_{3^{(0)}}  +  c^{d \, (1) *}_{2^{(0)}} c^{b \, (1)}_{2^{(0)}} ) \,  \hat g^Z_{L \,d^{(0)}d^{(0)}} 
 \cr
\noalign{\kern 5pt}
&\qquad
+(  c^{d\, (1)*}_{3^{(0)}} +   c^{d\, (2)*}_{3^{(0)}}) \, \delta \hat g^Z_{L \, b^{(0)}b^{(0)}} 
+ c^{d \, (1) *}_{2^{(0)}} c^{b \, (1)}_{2^{(0)}}  \, \delta \hat g^Z_{L \,s^{(0)}s^{(0)}}   + \cdots  \cr
\noalign{\kern 5pt}
&= (  c^{d\, (1)*}_{3^{(0)}} +   c^{d\, (2)*}_{3^{(0)}}) \, \delta \hat g^Z_{L \, b^{(0)}b^{(0)}} 
+ c^{d \, (1) *}_{2^{(0)}} c^{b \, (1)}_{2^{(0)}}  \, \delta \hat g^Z_{L \,s^{(0)}s^{(0)}}   + \cdots .
\label{ZcouplingB1}
\end{align}
In the last equality we have made use of the relations in Eqs.\ (\ref{downWave6}) and (\ref{downWave7}).
Remarkable cancellations among dominant terms are observed.
Thorough evaluation gives 
\begin{align}
&\frac{1}{\hat g_L^{W \nu_e e}} 
\begin{pmatrix} \hat  g^{Z\, ds}_L \cr \hat   g^{Z\, db}_L\cr  \hat   g^{Z\, sb}_L  \end{pmatrix}
= \begin{pmatrix} (-7.4 - 0.6 i) \times 10^{-7} \cr (-2.2 + 6.2 i) \times 10^{-7}  \cr -7.3 \times 10^{-6}  \end{pmatrix} ,  \cr
\noalign{\kern 5pt}
&\frac{1}{\hat g_L^{W \nu_e e}} 
\begin{pmatrix} \hat   g^{Z\, ds}_R \cr \hat g^{Z\, db}_R \cr  \hat  g^{Z\, sb}_R \end{pmatrix}
= \begin{pmatrix} - 1.0 \times 10^{-7}  \cr (1.4 + 0.9 i) \times 10^{-8}   \cr (-9.0 + 0.2 i) \times 10^{-8}  \end{pmatrix} .
\label{ZcouplingB2}
\end{align}
One sees that flavor-changing neutral currents (FCNCs) are naturally suppressed.
FCNCs induce the mixing of neutral mesons ($M = K, B_d, B_s$)   at the tree level,  
yielding  $\Delta m_M^\tree \sim (m_M f_M^2/3 m_Z^2)  |\hat  g^{Z \, d}_{M} |^2$
where $m_M$ and $f_M$ are the meson mass and
decay constant and $\hat  g^{Z \, d}_{M}$ is the relevant coupling in $\hat  g^{Z d}_{L}$ or
$\hat  g^{Z d}_{R}$. \cite{FCNC2020a, FCNCanalysis}
Making use of $(m_K, m_{B_d}, m_{B_s}) \sim  (0.498, 5.280, 5.367)\,$GeV and
$(f_K, f_{B_d}, f_{B_s}) \sim (0.156, 0.191, 0.274)\,$GeV, 
one finds, for $\theta_H = 0.10$ and $m_\KK = 13\,$TeV, 
$(\Delta m_K^\tree, \Delta m_{B_d}^\tree, \Delta m_{B_s}^\tree) \sim (2.7 \times 10^{-19}, \, 3.3 \times 10^{-18}, \,
8.7 \times 10^{-16})\,$GeV.
The experimental values are $m_{K_{L}^{0}} - m_{K_{S}^{0}} =  (3.484 \pm 0.006) \times 10^{-15}\,$GeV, 
$m_{B^{0}_{H}} - m_{B^{0}_{L}} =  (3.336 \pm 0.013) \times 10^{-13}\,$GeV and
$m_{B_{sH}^0} - m_{B_{sL}^0} = (1.1693 \pm 0.0004) \times 10^{-11}\,$GeV.\cite{PDG2024b}
The tree level contributions in GHU are much smaller than the  experimental values.

\begin{table}[tbh]
\renewcommand{\arraystretch}{1.3}
\begin{center}
\caption{Input  parameters and predictions.
$Z$ boson mass $m_Z$, Fermi constant $G_F$,   fine structure constant $\alpha_\EM$, Higgs boson mass $m_H$, 
and quark-lepton masses are input parameters of the model.
Predicted values for two cases $(\theta_H, m_\KK) = (0.1, 13\,{\rm TeV})$ and $(0.08, 16\,{\rm TeV})$ are
tabulated.   Brane interaction parameters and diagonal elements of the mass terms for $D^{\pm \alpha}$
are taken to be $(\mu_1, \mu_2, \mu_3)= (0.1, 0.1, 1.0)$ and $\tilde m_1=\tilde m_2= \tilde m_3 = 0.8$
in both cases. 
In the last column the values in the SM global fit with $\sin^2 \theta_W^\SM = 0.2312$ are listed.
The angles $\alpha, \beta, \gamma$ of the unitarity triangle are defined in Eq.\ (\ref{unitarity1}).
The evaluated value of $m_W$ is from ref.\ \cite{Wmass2023}.
}
\vskip 10pt
\begin{tabular}{cccc}
\hline \hline 
Input & Case 1 & Case 2 &SM\\
$(\theta_H, m_\KK)$ &$(0.1, 13\,{\rm TeV})$ &$(0.08, 16\,{\rm TeV})$ & \\
$\tilde m_{12} = \tilde m_{21}^*$ & $1.18 \times 10^{-2}$ &  $1.20\times 10^{-2}$&  \\
$\tilde m_{23} = \tilde m_{32}^*$ & $8.10 \times 10^{-5}$ &$7.98 \times 10^{-5}$ &  \\
$\tilde m_{13} = \tilde m_{31}^*$ & $(1.52 - 3.74 i) \times 10^{-6}$ &$(1.45 - 3.62 i) \times 10^{-6}$ &  \\
\hline
Output & &\\
$k$ & $1.586 \times 10^{15}\,$GeV &  $8.997 \times 10^{14}\,$GeV & --- \\
$\hat g_L^{W \nu_e e}$ &$0.997647$  &  $0.998497$  &1 \\
$\alpha$ &$ 89.97 {\,}^\circ$ & $ 89.90 {\,}^\circ$ & $ 91.6 {\,}^\circ$ \\
$\sin 2\beta$ &$0.7071$ &$0.6999$  &0.713 \\
$\gamma$ &$ 67.53 {\,}^\circ$ &  $ 67.89 {\,}^\circ$&  $ 65.7 {\,}^\circ$\\
$\myfrac{1}{\hat g_L^{W \nu_e e}} 
\begin{pmatrix} \hat  g^{Z\, dd}_L \cr \hat   g^{Z\, dd}_R  \end{pmatrix}$ &
$\begin{pmatrix} -0.42293 \cr 0.07697  \end{pmatrix}$ & $\begin{pmatrix} -0.42289 \cr 0.07701  \end{pmatrix}$
&$\begin{pmatrix} -0.4229 \cr 0.0771  \end{pmatrix}$\\
$\myfrac{1}{\hat g_L^{W \nu_e e}} 
\begin{pmatrix} \hat  g^{Z\, ds}_L \cr \hat   g^{Z\, ds}_R  \end{pmatrix}$ &
$\begin{pmatrix} (-7.4 - 0.6 i) \times 10^{-7} \cr -1.0 \times 10^{-7}  \end{pmatrix}$ 
& $\begin{pmatrix} (-5.6 - 0.5 i) \times 10^{-7} \cr - 0.7 \times 10^{-7}  \end{pmatrix}$
&0 \\
$m_W$ &$80.396\,$GeV & $80.381\,$GeV &$80.354\,$GeV\\
\hline \hline
\end{tabular}
\label{Tab:parameter}
\end{center}
\end{table}

\section{Summary} 

In this paper we have examined the flavor problem in the quark sector in the GUT inspired 
$SO(5) \times U(1)_X \times SU(3)_C$ GHU in the RS space. 
Masses of down-type quarks are split from those of up-type quarks by $SO(5)$ singlet fermion multiplets 
 $\Psi_{({\bf 3,1})}^{\pm \alpha}$ which naturally appear in gauge-Higgs grand unification models.  
Fields $\Psi_{({\bf 3,1})}^{\pm \alpha}$ are important to suppress   $W$ couplings of right-handed components
of quarks almost entirely as well.
General masses of those  $SO(5)$ singlet fermion multiplets
induce the mixing in the $W$ couplings of quarks.  The observed CKM matrix in the $W$ couplings is
reproduced within experimental errors.
Furthermore we have evaluated flavor-changing $Z$ couplings at the tree level to find to be extremely small.
FCNCs are naturally suppressed as a result of remarkable cancellations among the coefficient
functions $c^{\beta \, (j)}_{\alpha^{(n)}}$ manifested by the relations in Eqs.\  (\ref{downWave6}) and (\ref{downWave7}).

Acceptable phenomenological values for the $W$ and $Z$ couplings of quarks are obtained 
for $\theta_H \lesssim 0.1$.  As has been shown in ref.\ \cite{Wmass2023}, the value of the $W$ boson mass, $m_W$,
sensitively depends on $\theta_H$.  Precise measurements of $m_W$ will help to narrow the allowed 
range of $\theta_H$.

The mixing in the neutrino sector in GHU needs to be clarified as a next problem.  Majorana masses on the UV brane
in the GUT inspired $SO(5) \times U(1)_X \times SU(3)_C$  GHU are expected to play a  crucial role in this regard.

\section*{Acknowledgment}

This work was supported in part by the Japan Society for the Promotion of Science (JSPS) 
KAKENHI Grant No. JP21H05182 (N.Y.).

\vskip 1.cm 

\appendix

\section{Basis functions} 

We  summarize basis functions used for wave functions of gauge and fermion fields.
For gauge fields we introduce
\begin{align}
 F_{\alpha, \beta}(u, v) &\equiv J_\alpha(u) Y_\beta(v) - Y_\alpha(u) J_\beta(v) ~, \cr
\noalign{\kern 5pt}
 C(z; \lambda) &= \frac{\pi}{2} \lambda z z_L F_{1,0}(\lambda z, \lambda z_L) ~,  \cr
 S(z; \lambda,) &= -\frac{\pi}{2} \lambda  z F_{1,1}(\lambda z, \lambda z_L) ~, \cr
 C^\prime (z; \lambda) &= \frac{\pi}{2} \lambda^2 z z_L F_{0,0}(\lambda z, \lambda z_L) ~,  \cr
S^\prime (z; \lambda) &= -\frac{\pi}{2} \lambda^2 z  F_{0,1}(\lambda z, \lambda z_L)~, 
\label{functionA1}
\end{align}
where $J_\alpha (u)$ and $Y_\alpha (u)$ are Bessel functions of  the first and second kind.
They satisfy
\begin{align}
&- z \frac{d}{dz} \frac{1}{z} \frac{d}{dz} \begin{pmatrix} C \cr S \end{pmatrix} 
= \lambda^{2} \begin{pmatrix} C \cr S \end{pmatrix} ~,  
\label{relationA1}
\end{align}
with the boundary conditions $C = z_L$, $C'  =S = 0 $ and $S' = \lambda$ at $z=z_L$, 
and a relation $CS' - S C' = \lambda z$ holds.

For fermion fields with a bulk mass parameter $c$, we define 
\begin{align}
\begin{pmatrix} C_L \cr S_L \end{pmatrix} (z; \lambda,c)
&= \pm \frac{\pi}{2} \lambda \sqrt{z z_L} F_{c+\frac12, c\mp\frac12}(\lambda z, \lambda z_L) ~, \cr
\begin{pmatrix} C_R \cr S_R \end{pmatrix} (z; \lambda,c)
&= \mp \frac{\pi}{2} \lambda \sqrt{z z_L} F_{c- \frac12, c\pm\frac12}(\lambda z, \lambda z_L) ~, \cr
\begin{pmatrix} \check S_L \cr \check C_R \end{pmatrix} (z; \lambda,c)
&=\frac{C_L (1; \lambda, c)}{S_L (1; \lambda, c)} \begin{pmatrix} S_L \cr C_R \end{pmatrix} (z; \lambda,c) ~.
\label{functionA2}
\end{align}
These functions satisfy 
\begin{align}
&D_{+} (c) \begin{pmatrix} C_{L} \cr S_{L} \end{pmatrix} = \lambda  \begin{pmatrix} S_{R} \cr C_{R} \end{pmatrix}, \cr
\noalign{\kern 5pt}
&D_{-} (c) \begin{pmatrix} S_{R} \cr C_{R} \end{pmatrix} = \lambda  \begin{pmatrix} C_{L} \cr S_{L} \end{pmatrix}, ~~
D_{\pm} (c) = \pm \frac{d}{dz} + \frac{c}{z} ~, 
\label{relationA2}
\end{align}
with the boundary conditions $C_{R/L} =1$, $D_-(c) C_R = D_+(c) C_L =S_{R/L} = 0$ at $z=z_{L} $, and 
$C_L C_R - S_L S_R=1$.

For fermion fields with a vector-like mass $m = k \tilde m$,  mode functions are expressed
in terms of
\begin{align}
{\cal C}_{L/R\, 1}(z; \lambda, c, \tilde m) &= C_{L/R} (z; \lambda, c+\tilde{m})+C_{L/R} (z; \lambda, c-\tilde{m}) ~, \cr
{\cal S}_{L/R\, 1}(z; \lambda, c, \tilde m) &= S_{L/R} (z; \lambda, c+\tilde{m})+S_{L/R} (z; \lambda,c-\tilde{m}) ~, \cr
{\cal C}_{L/R\, 2}(z; \lambda, c, \tilde m) &= S_{L/R} (z; \lambda, c+\tilde{m})-S_{L/R} (z; \lambda, c-\tilde{m}) ~, \cr
{\cal S}_{L/R\, 2}(z; \lambda, c, \tilde m) &= C_{L/R} (z; \lambda, c+\tilde{m})-C_{L/R} (z; \lambda, c-\tilde{m}) ~.
\label{MassiveFermion1}
\end{align}
These functions satisfy
\begin{align}
&D_+ (c) \begin{pmatrix} {\cal C}_{L1} \cr {\cal S}_{L2} \end{pmatrix}
+ \frac{\tilde m}{z} \begin{pmatrix} {\cal S}_{L2} \cr {\cal C}_{L1} \end{pmatrix} 
= \lambda \begin{pmatrix} {\cal S}_{R1} \cr {\cal C}_{R2} \end{pmatrix}  , \cr
\noalign{\kern 5pt}
&D_- (c) \begin{pmatrix} {\cal S}_{R1} \cr {\cal C}_{R2} \end{pmatrix}
+  \frac{\tilde m}{z} \begin{pmatrix} {\cal C}_{R2} \cr {\cal S}_{R1} \end{pmatrix}
= \lambda \begin{pmatrix} {\cal C}_{L1} \cr {\cal S}_{L2} \end{pmatrix} , \cr
\noalign{\kern 5pt}
&D_+ (c) \begin{pmatrix} {\cal S}_{L1} \cr {\cal C}_{L2} \end{pmatrix}
+  \frac{\tilde m}{z} \begin{pmatrix} {\cal C}_{L2} \cr {\cal S}_{L1} \end{pmatrix}
= \lambda \begin{pmatrix} {\cal C}_{R1} \cr {\cal S}_{R2} \end{pmatrix} , \cr
\noalign{\kern 5pt}
&D_- (c) \begin{pmatrix} {\cal C}_{R1} \cr {\cal S}_{R2} \end{pmatrix}
+  \frac{\tilde m}{z} \begin{pmatrix} {\cal S}_{R2} \cr {\cal C}_{R1} \end{pmatrix}
= \lambda \begin{pmatrix} {\cal S}_{L1} \cr {\cal C}_{L2} \end{pmatrix} , 
\label{MassiveFermion2}
\end{align}
and ${\cal S}_{Rj} = {\cal S}_{Lj} = D_-(c)\,  {\cal C}_{Rj} = D_+(c) \,  {\cal C}_{Lj} = 0$ at $z=z_L$.

In evaluating $W_{\alpha^{(n)}, \beta^{(\ell)}}$ in Eq.\ (\ref{Velement1}), behavior of wave functions 
near the UV brane at $z=1$ becomes important.  In numerical evaluation the identities
\begin{align}
&S_L (1; \lambda,c) C_L (z; \lambda,c) - C_L (1; \lambda,c) S_L (z; \lambda,c)
= - \frac{\pi}{2} \lambda \sqrt{z} \, F_{c+\onehalf, c+\onehalf} (\lambda, \lambda z) , \cr
\noalign{\kern 5pt}
&S_L (1; \lambda,c) S_R (z; \lambda,c) - C_L (1; \lambda,c) C_R (z; \lambda,c)
=  - \frac{\pi}{2} \lambda \sqrt{z} \, F_{c+\onehalf, c-\onehalf} (\lambda, \lambda z)  ,
\label{relationA3}
\end{align}
are useful.

\section{$SO(5)$ generators} 

Generators of $SO(5)$ in the spinor representation are given by
\begin{align}
&T^{a_L} = \onehalf \sigma^a \otimes \begin{pmatrix} 1&0 \cr 0&0 \end{pmatrix}  ,~~
T^{a_R} = \onehalf \sigma^a \otimes \begin{pmatrix} 0&0 \cr 0&1 \end{pmatrix} , \cr
&\hat T^a = - \frac{1}{2 \sqrt{2}} ~ \sigma^a \otimes \sigma^2 ,~~
\hat T^4 =  \frac{1}{2 \sqrt{2}} ~ \sigma^0 \otimes \sigma^1 ,  \quad (a=1,2,3). 
\label{so5generator}
\end{align}
In this representation 
\begin{align}
& \Psi_{\bf (3,4)}^{\alpha=1} = \begin{pmatrix} u \cr d \cr u' \cr d' \end{pmatrix} , \quad
 \hat \Phi_{({\bf 1}, {\bf 4}) } =  \begin{pmatrix} \hat \Phi_{[{\bf 2}, {\bf 1}]} \cr \hat \Phi_{[{\bf 1}, {\bf 2}]} \end{pmatrix} .
\end{align}

\section{Coefficients $c^d_{\alpha ^{(n)}}$, $c^s_{\alpha ^{(n)}}$ and $c^b_{\alpha ^{(n)}}$} 

Note that $\{ 1^{(n)} \} = \{ d^{(n)} , n \ge 0 \}  + \{ D_d^{(n)} , n \ge 1\} $ etc. 
Coefficients $c^d_{\alpha ^{(n)}}$, $c^s_{\alpha ^{(n)}}$ and $c^b_{\alpha ^{(n)}}$ in Eq.\ (\ref{downWave4}) 
for $\theta_H=0.1$ and $m_\KK = 13\,$TeV with mass parameters in Eq.\ (\ref{tilmass1}) are,
to the first order in perturbation theory, given by $c^{d \, (1)}_{1^{(n)}} =c^{s \, (1)}_{2^{(n)}} = c^{b \, (1)}_{3^{(n)}} = 0$ and 
\begin{align}
&c^{d \, (1)}_{s^{(n)}} = 
\begin{pmatrix} - 0.2194 \cr 10^{-5} \cr   10^{-6} \cr  10^{-5} \end{pmatrix}, ~
c^{d \, (1)}_{D_s^{(n)}} = \begin{pmatrix} 0.0116 \cr 0.0058 \cr - 0.0055 \cr - 0.0042 \end{pmatrix}, ~
c^{d \, (1)}_{b^{(n)}} =   \begin{pmatrix} -0.0014 - 0.0034 i\cr 10^{-5} \cr  10^{-6} \cr 10^{-6} \end{pmatrix}, \cr
\noalign{\kern 5pt}
&c^{s \, (1)}_{d^{(n)}} =  \begin{pmatrix} 0.2194 \cr 10^{-6}\cr 10^{-8} \cr 10^{-6} \end{pmatrix}, ~
c^{s \, (1)}_{D_d^{(n)}} = \begin{pmatrix} 0.0129 \cr 0.0063 \cr - 0.0058 \cr - 0.0044 \end{pmatrix}, ~
c^{s \, (1)}_{b^{(n)}} =   \begin{pmatrix} -0.0408 \cr 0.0005 \cr  10^{-5} \cr - 0.0002 \end{pmatrix}, \cr
\noalign{\kern 5pt}
&c^{b \, (1)}_{d^{(n)}} =  \begin{pmatrix} 0.0014 - 0.0034 i\cr 10^{-10} \cr  10^{-12} \cr 10^{-11} \end{pmatrix}, ~
c^{b \, (1)}_{s^{(n)}} =  \begin{pmatrix} 0.0408 \cr10^{-8} \cr  10^{-9} \cr 10^{-8} \end{pmatrix}, ~
c^{b \, (1)}_{D_s^{(n)}} =  \begin{pmatrix} 0.0001 \cr10^{-5} \cr  10^{-5} \cr 10^{-5} \end{pmatrix} .
\label{coeff1}
\end{align}
Here $c^{d \, (1)}_{s^{(n)}} $ ($ n =0,1,2,3$) and $c^{d \, (1)}_{D_s^{(n)}} $ ($n=1,2,3,4$) are shown, and  $10^{-6}$ means
$O(10^{-6})$.  Suppressed coefficients $c^{d \, (1)}_{D_b^{(n)}} $ etc.\ are all $O(10^{-5})$ or less.

To the second order in perturbation theory,  relevant coefficients are
\begin{align}
&c^{d \, (2)}_{d^{(0)}} = - 0.0242 ~, ~ c^{d \, (2)}_{s^{(0)}} = 0.0030 + 0.0075 i ~, ~
c^{d \, (2)}_{b^{(0)}} = 0.00878~, \cr
\noalign{\kern 5pt}
&c^{s \, (2)}_{s^{(0)}} = - 0.0250 ~,~ c^{s \, (2)}_{d^{(0)}} = - 0.0031 + 0.0076 i ~, ~
c^{s \, (2)}_{b^{(0)}} = - 0.0003 - 0.0008 i ~,  \cr
\noalign{\kern 5pt}
&c^{b \, (2)}_{b^{(0)}} = - 0.00084 ~, ~ c^{b \, (2)}_{d^{(0)}} = 0.00016 ~, ~
c^{b \, (2)}_{s^{(0)}} = (0.5 - 1.3 i ) \times 10^{-5} ~.
\end{align}

\section{$W$ and $Z$ couplings before mixing} 

Left-handed $W$  couplings before mixing 
defined in Eq.\ (\ref{Wcoupling2}), are for $\theta_H=0.1$, $m_\KK = 13\,$TeV and 
 $\sin^2 \theta_W^0 = 0.23035$ given by
\begin{align}
&\hat g^W_{L\,u^{(0)} d^{(n)}} = \begin{pmatrix} 0.997645 \cr -0.02490 \cr 1.9 \times 10^{-5} \cr - 2.83 \times 10^{-3} \end{pmatrix},~~
\hat g^W_{L\,u^{(0)}  D_d^{(n)}} = \begin{pmatrix} 4 \times 10^{-7} \cr - 1 \times 10^{-7} \cr  7 \times 10^{-8} \cr   -2 \times 10^{-8} \end{pmatrix},\cr
\noalign{\kern 5pt}
&\hat g^W_{L\,c^{(0)} s^{(n)}} = \begin{pmatrix} 0.997643 \cr -0.02627 \cr 2.1 \times 10^{-5} \cr - 2.67 \times 10^{-3} \end{pmatrix}, ~~
\hat g^W_{L\,c^{(0)}  D_s^{(n)}} = \begin{pmatrix} 1.0 \times 10^{-5} \cr - 2 \times 10^{-6} \cr  1 \times 10^{-6} \cr   -3 \times 10^{-7} \end{pmatrix},\cr
\noalign{\kern 5pt}
&\hat g^W_{L\,t^{(0)} b^{(n)}} = \begin{pmatrix} 0.997969 \cr 0.03174 \cr -1.45 \times 10^{-4} \cr  1.98 \times 10^{-3} \end{pmatrix}, ~~
\hat g^W_{L\,t^{(0)}  D_b^{(n)}} = \begin{pmatrix} - 1.52 \times 10^{-3} \cr 7.7 \times 10^{-5} \cr -3.6 \times 10^{-5} \cr   0.6 \times 10^{-5} \end{pmatrix},
\end{align}
It is seen that most of the couplings of KK modes other than $\hat g^W_{L\,u^{(0)} d^{(1)}}$, $\hat g^W_{L\,c^{(0)} s^{(1)}}$
and $\hat g^W_{L\,t^{(0)} b^{(1)}}$ are very small.

$Z$ couplings in the down-type quark sector before mixing $\hat g^{Z d(nm) }_{L/R}$, $\hat g^{Z s(nm) }_{L/R}$ and 
$\hat g^{Zb(nm) }_{L/R}$ defined in Eq.\ (\ref{Zcoupling2b}) are given by
\begin{align}
&\hat g^{Z}_{L\,  d^{(n)} d^{(m)}} = 
\begin{pmatrix} -0.4230 & 0.0125 & 10^{-5} & 0.0014  & 10^{-7}\cr
0.0125 & 0.0757 & - 0.0142 & 10^{-6} & - 0.0003 \cr
10^{-5} & -0.0142 & -0.4230 & 0.0123 & 10^{-5} \cr
0.0014 & 10^{-6} & 0.0123 & 0.0757 & -0.0135 \cr
10^{-7} & -0.0003 & 10^{-5} &-0.0135 & - 0.4230 \end{pmatrix}, \cr
\noalign{\kern 5pt}
&\hat g^{Z}_{L\,  s^{(n)} s^{(m)}} = 
\begin{pmatrix} -0.4230 & 0.0132 & 10^{-5} & 0.0013  & 10^{-6}\cr
0.0132 & 0.0757 & - 0.0142 & 10^{-6} & - 0.0003 \cr
10^{-5} & -0.0142 & -0.4230 & 0.0125 & 10^{-5} \cr
0.0013 & 10^{-6} & 0.0125 & 0.0757 & -0.0135 \cr
10^{-6} & -0.0003 & 10^{-5} &-0.0135 & - 0.4230 \end{pmatrix}, \cr
\noalign{\kern 5pt}
&\hat g^{Z}_{L\,  b^{(n)} b^{(m)}} = 
\begin{pmatrix} -0.4230 & -0.0159 & 10^{-5} &- 0.0010  & 10^{-6}\cr
-0.0159 & 0.0757 & 0.0135 & 10^{-6} &  0.0004 \cr
10^{-5} & 0.0135 & -0.4230 & - 0.0123 & 10^{-5} \cr
-0.0010 & 10^{-6} & -0.0123 & 0.0757 & 0.0132 \cr
10^{-6} & 0.0004 & 10^{-5} & 0.0132 & - 0.4230 \end{pmatrix},
\label{ZcouplingList1}
\end{align}
and
\begin{align}
&\hat g^{Z}_{R\,  d^{(n)} d^{(m)}} = \begin{pmatrix} 0.0770 & 10^{-8} & 10^{-8} & 10^{-10}  & 10^{-9}\cr
10^{-8}& 0.0757 & - 0.0121 & 10^{-6} & 10^{-6} \cr
10^{-8} & -0.0121 & -0.4230 & 0.0113 & 10^{-5} \cr
10^{-10} & 10^{-6} & 0.0113 & 0.0757 & -0.0129 \cr
10^{-9} & 10^{-6} & 10^{-5} &-0.0129 & - 0.4230 \end{pmatrix}, \cr
\noalign{\kern 5pt}
&\hat g^{Z}_{R\,  s^{(n)} s^{(m)}} = \begin{pmatrix} 0.0770 & 10^{-7} & 10^{-7} & 10^{-10}  & 10^{-7}\cr
10^{-7}& 0.0757 & - 0.0123 & 10^{-6} & - 0.0001 \cr
10^{-7} & -0.0123 & -0.4230 & 0.0115 &  10^{-5} \cr
10^{-10} & 10^{-6} & 0.0115 & 0.0757 & -0.0130 \cr
10^{-7} & -0.0001 & 10^{-5} &-0.0130 & - 0.4230 \end{pmatrix}, \cr
\noalign{\kern 5pt}
&\hat g^{Z}_{R\,  b^{(n)}b^{(m)}} = \begin{pmatrix} 0.0770 & 10^{-5} & 10^{-5} & 10^{-6}  & 10^{-6}\cr
10^{-5}& 0.0757 &  0.0129 & 10^{-6} &  0.0004 \cr
10^{-5} &  0.0129 & -0.4230 & - 0.0123 &  10^{-5} \cr
10^{-6} & 10^{-6} & -0.0123 & 0.0757 & 0.0130 \cr
10^{-6} & 0.0004 & 10^{-5} & 0.0130 & - 0.4230 \end{pmatrix} .
\label{ZcouplingList2}
\end{align}
Couplings of $D_{d,s,b}^{(n)}$ modes are small except for diagonal elements
\begin{align}
&\hat g^{Z}_{L/R\,  D_d^{(n)} D_d^{(n)}},  ~ \hat g^{Z}_{L/R\,  D_s^{(n)} D_s^{(n)}}, ~ \hat g^{Z}_{L/R\,  D_s^{(n)} D_s^{(n)}} \cr
\noalign{\kern 5pt}
&\qquad
\sim - Q_d \frac{\sin^2 \theta_W^0}{ \hat g^W_{L\, \nu_e e}}  \sim 0.0770 ~.
\label{ZcouplingList3}
\end{align}


\def\jnl#1#2#3#4{{#1}{\bf #2},  #3 (#4)}

\def\Zphys{{\em Z.\ Phys.} }
\def\jssc{{\em J.\ Solid State Chem.\ }}
\def\jpsJ{{\em J.\ Phys.\ Soc.\ Japan }}
\def\ptps{{\em Prog.\ Theoret.\ Phys.\ Suppl.\ }}
\def\PTP{{\em Prog.\ Theoret.\ Phys.\  }}
\def\PTEP{{\em Prog.\ Theoret.\ Exp.\  Phys.\  }}
\def\JMP{{\em J. Math.\ Phys.} }
\def\NPB{{\em Nucl.\ Phys.} B}
\def\NP{{\em Nucl.\ Phys.} }
\def\PLB{{\it Phys.\ Lett.} B}
\def\PL{{\em Phys.\ Lett.} }
\def\PRL{\em Phys.\ Rev.\ Lett. }
\def\PRB{{\em Phys.\ Rev.} B}
\def\PRD{{\em Phys.\ Rev.} D}
\def\PRe{{\em Phys.\ Rep.} }
\def\AP{{\em Ann.\ Phys.\ (N.Y.)} }
\def\RMP{{\em Rev.\ Mod.\ Phys.} }
\def\ZPC{{\em Z.\ Phys.} C}
\def\SCI{{\em Science} }
\def\CMP{\em Comm.\ Math.\ Phys. }
\def\MPLA{{\em Mod.\ Phys.\ Lett.} A}
\def\IJMPA{{\em Int.\ J.\ Mod.\ Phys.} A}
\def\IJMPB{{\em Int.\ J.\ Mod.\ Phys.} B}
\def\EPJC{{\em Eur.\ Phys.\ J.} C}
\def\EPJP{{\em Eur.\ Phys.\ J.} Plus}
\def\PR{{\em Phys.\ Rev.} }
\def\JHEP{{\em JHEP} }
\def\JCAP{{\em JCAP} }
\def\cmp{{\em Com.\ Math.\ Phys.}}
\def\JPA{{\em J.\  Phys.} A}
\def\JPG{{\em J.\  Phys.} G}
\def\NJP{{\em New.\ J.\  Phys.} }
\def\CQG{\em Class.\ Quant.\ Grav. }
\def\ATMP{{\em Adv.\ Theoret.\ Math.\ Phys.} }
\def\ibid{{\em ibid.} }
\def\ChP{{\em Chin.Phys.}C}
\def\NCA{{\it Nuovo Cim.} A}

\renewenvironment{thebibliography}[1]
         {\begin{list}{[$\,$\arabic{enumi}$\,$]}  
         {\usecounter{enumi}\setlength{\parsep}{0pt}
          \setlength{\itemsep}{0pt}  \renewcommand{\baselinestretch}{1.2}
          \settowidth
         {\labelwidth}{#1 ~ ~}\sloppy}}{\end{list}}

 \vskip 1.cm

\leftline{\Large \bf References}



\begin{thebibliography}{99}


\bibitem{Hosotani1983}
Y.\ Hosotani, 
{\it ``Dynamical mass generation by compact extra dimensions''}, 
\jnl{\PLB}{126}{309}{1983}.

\bibitem{Davies1988}
A.~T.~Davies and A.~McLachlan,
{\it ``Gauge group breaking by Wilson loops''},
\jnl{\PLB}{200}{305}{1988}.

\bibitem{Hosotani1989}
Y.\ Hosotani,  
{\it ``Dynamics of nonintegrable phases and gauge symmetry breaking''}, 
\jnl{\AP}{190}{233}{1989}.

\bibitem{Davies1989}
A.~T.~Davies and A.~McLachlan,
{\it ``Congruency class effects in the Hosotani model''},
\jnl{\NPB}{317}{237}{1989}.

\bibitem{HetrickHo1989}
J.E.~Hetrick and C-L.~Ho,
{\it ``Dynamical symmetry breaking from toroidal compactification''},
\jnl{\PRD}{40}{4085}{1989}.

\bibitem{McLachlan1990}
A.~McLachlan,
{\it ``Flux-breaking in space-times with toroidal compactification''},
\jnl{\NPB}{338}{188}{1990}.

\bibitem{Hatanaka1998}
H.\ Hatanaka, T.\ Inami, C.S.\ Lim,
{\it ``The gauge hierarchy problem and higher dimensional gauge theories''}, 
\jnl{\MPLA}{13}{2601}{1998}.

\bibitem{Hatanaka1999}
H.\ Hatanaka,
{\it ``Matter representations and gauge symmetry breaking via compactified space''}, 
\jnl{\PTP}{102}{407}{1999}.

\bibitem{Antoniadis2001}
I.~Antoniadis, K.~Benakli and M.~Quiros,
{\it ``Finite Higgs mass without supersymmetry''},
\jnl{\NJP}{3}{20}{2001}.

\bibitem{Takenaga2002}
K.~Takenaga,
{\it ``Gauge symmetry breaking through the Hosotani mechanism in softly broken supersymmetric QCD''},
\jnl{\PRD}{66}{085009}{2002}.


\bibitem{Kubo2002}
M.\ Kubo, C.S.\ Lim and H.\ Yamashita, 
{\it ``The Hosotani mechanism in bulk gauge theories with an orbifold extra space $S^1/Z_2$''}, 
\jnl{\MPLA}{17}{2249}{2002}.




\bibitem{BurdmanNomura2003}
G.~Burdman and Y.~Nomura,
{\it ``Unification of Higgs and gauge fields in five dimensions''},
\jnl{\NPB}{656}{3}{2003}.

\bibitem{Csaki2003}
C.~Csaki, C.~Grojean and H.~Murayama,
{\it ``Standard model Higgs from higher dimensional gauge fields''},
\jnl{\PRD}{67}{085012}{2003}.


\bibitem{Scrucca2003}
C.A.~Scrucca, M.~Serone,  L.~Silvestrini,
{\it ``Electroweak symmetry breaking and fermion masses from extra dimensions''}, 
\jnl{\NPB}{669}{128}{2003}.

\bibitem{ACP2005}
K.~Agashe, R.~Contino and A.~Pomarol,
{\it ``The minimal composite Higgs model''}, 
\jnl{\NPB}{719}{165}{2005}.

\bibitem{Cacciapaglia2006}
G.~Cacciapaglia, C.~Csaki, S.C.~Park,
{\it ``Fully radiative electroweak symmetry breaking''}, 
\jnl{\JHEP}{0603}{099}{2006}.

\bibitem{Medina2007}
A.~D.~Medina, N.~R.~Shah and C.~E.~M.~Wagner, 
{\it ``Gauge-Higgs unification and radiative electroweak symmetry breaking in warped 
extra dimensions''}, 
\jnl{\PRD}{76}{095010}{2007}.

\bibitem{HOOS2008} 
Y.~Hosotani, K.~Oda, T.~Ohnuma and Y.~Sakamura,
{\it ``Dynamical electroweak symmetry breaking in $SO(5) \times U(1)$ gauge-Higgs 
unification with top and bottom quarks''}, 
\jnl{\PRD}{78}{096002}{2008}; 
{\it Erratum}-\jnl{\ibid}{{\rm D}79}{079902}{2009}.


\bibitem{Serone2010}
M.~Serone,
{\it ``Holographic methods and gauge-Higgs unification in flat extra dimensions''},
\jnl{\NJP}{12}{075013}{2010}.


\bibitem{FHHOS2013} 
 S.~Funatsu, H.~Hatanaka, Y.~Hosotani, Y.~Orikasa and T.~Shimotani,
{\it ``Novel universality and Higgs decay $H\to \gamma\gamma, gg$ in the 
$SO(5) \times U(1)$ gauge-Higgs unification''}, 
\jnl{\PLB}{722}{94}{2013}.


\bibitem{Yoon2018b}
J.~Yoon and M.~E. Peskin, 
{\it ``Dissection of an $SO(5) \times U(1)$ gauge-Higgs  unification model''},
\jnl{\PRD}{100}{015001}{2019}. 


\bibitem{GUTinspired2019a}
 S.~Funatsu, H.~Hatanaka, Y.~Hosotani, Y.~Orikasa and N.~Yamatsu,
{\it ``GUT inspired $SO(5)\times U(1) \times SU(3)$ gauge-Higgs unification''}, 
\jnl{\PRD}{99}{095010}{2019}.   


\bibitem{FCNC2020a}
 S.~Funatsu, H.~Hatanaka, Y.~Hosotani, Y.~Orikasa and N.~Yamatsu,
{\it ``CKM matrix and FCNC suppression in $SO(5)\times U(1) \times SU(3)$ gauge-Higgs unification''}, 
\jnl{\PRD}{101}{055016}{2020}.   

\bibitem{GUTinspired2020b}
 S.~Funatsu, H.~Hatanaka, Y.~Hosotani, Y.~Orikasa and N.~Yamatsu,
{\it ``Effective potential and universality in GUT-inspired gauge-Higgs unification''}, 
\jnl{\PRD}{102}{015005}{2020}.   



\bibitem{RS1}
L.\ Randall and R.\ Sundrum,
{\it ``A large mass hierarchy from a small extra dimension''}, 
\jnl{\PRL}{83}{3370}{1999}.


\bibitem{SO11GHGU}
Y.~Hosotani and N.~Yamatsu, 
{\it ``Gauge-Higgs grand unification''}, 
\jnl{\PTEP}{2015}{111B01}{2015}, arXiv:1504.03817 [hep-ph]; 
A.~Furui, Y.~Hosotani, and N.~Yamatsu, 
{\it ``Toward realistic gauge-Higgs grand  unification''}, 
\jnl{\PTEP}{2016}{093B01}{2016}, arXiv:1606.07222 [hep-ph].



\bibitem{Funatsu2017a}
S.~Funatsu, H.~Hatanaka, Y.~Hosotani, and Y.~Orikasa, 
{\it ``Distinct signals of  the gauge-Higgs unification in $e^+e^-$ collider experiments''}, 
\jnl{\PLB}{775}{297}{2017}.   

\bibitem{Yoon2018a}
J.~Yoon and M.~E. Peskin, 
{\it ``Fermion pair production in $SO(5) \times U(1)$  gauge-Higgs unification models''}, 
arXiv:1811.07877 [hep-ph].

\bibitem{Funatsu2019a}
S.~Funatsu, 
{\it ``Forward-backward asymmetry in the gauge-Higgs unification at the International Linear Collider''},
\jnl{\EPJC}{79}{854}{2019}.   


\bibitem{GUTinspired2020c}
 S.~Funatsu, H.~Hatanaka, Y.~Hosotani, Y.~Orikasa and N.~Yamatsu,
{\it ``Fermion pair production at $e^- e^+$ linear collider experiments
in GUT inspired gauge-Higgs unification''}, 
\jnl{\PRD}{102}{015029}{2020}.

 

\bibitem{Irles2024}
A.~Irles, J.P.~M\'{a}rquez, R.~P\"{o}schl, F.~Richard, A.~Saibel, H.~Yamamoto and N.~Yamatsu,
{\it ``Probing gauge-Higgs unification models at the ILC with quark-antiquark forward-backward asymmetry 
at center-of-mass energies above the $Z$ mass''},
\jnl{\EPJC}{84}{537}{2024}.  


 \bibitem{Funatsu2023a}
 S.~Funatsu, H.~Hatanaka, Y.~Orikasa, and N.~Yamatsu, 
 {\it ``Single Higgs boson production at electron-positron colliders in gauge-Higgs unification"},
 \jnl{\PRD}{107}{075030}{2023}.

\bibitem{Yamatsu2023}
N.~Yamatsu,  S.~Funatsu, H.~Hatanaka, Y.~Hosotani,  and Y.~Orikasa
{\it ``$W$  and $Z$ boson pair production at electron-positron colliders  in gauge-Higgs unification''}, 
\jnl{\PRD}{108}{115014}{2023}.


\bibitem{Wmass2023}
 Y.~Hosotani,  S.~Funatsu, H.~Hatanaka,Y.~Orikasa and N.~Yamatsu,
{\it ``$W$ boson mass in gauge-Higgs unification''}, 
\jnl{\PRD}{108}{115036}{2023}.

\bibitem{CDF2022}
T.~Aaltonen et al.\ (CDF Collaboration),
{\it ``High-precision measurement of the $W$ boson  mass with the CDF II detector''},
\jnl{\SCI}{376}{170}{2022}.

\bibitem{ATLAS2023}
The ATLAS Collaboration,
{\it ``Improved $W$ boson mass measurement using $\sqrt{s} = 7\,$TeV proton-proton collisions with the ATLAS detector''},
Report No. ATLAS-CONF-2023-004.

\bibitem{CMS2024}
The CMS Collaboration,
{\it ``High-precision measurement of the  $W$ boson mass  with the CMS experiment at the LHC''},
Report No. CMS-SMP-23-002. (arXiv:2412.13872 [hep-ex])



\bibitem{LimMaru2007}
C.S.~Lim and N.~Maru,
{\it ``Towards a realistic grand gauge-Higgs unification''},
\jnl{\PLB}{653}{320}{2007}.   

\bibitem{Kojima2011}
K.~Kojima, K.~Takenaga and T.~Yamashita,
{\it ``Grand gauge-Higgs unification''},
\jnl{\PRD}{84}{051701(R)}{2011};
{\it ``Gauge symmetry breaking patterns in an $SU(5)$ grand gauge-Higgs unification model''},
\jnl{\PRD}{95}{015021}{2017}.

\bibitem{HosotaniYamatsu2018}
Y.~Hosotani and N.~Yamatsu, 
{\it ``Gauge-Higgs seesaw mechanism in 6-dimensional grand unification''}, 
 \jnl{\PTEP}{2017}{091B01}{2017},  (arXiv:1706.03503 [hep-ph]); 
{\it ``Electroweak symmetry breaking and mass spectra
  in six-dimensional gauge-Higgs grand unification''},
 \jnl{\PTEP}{2018}{023B05}{2018},  (arXiv:1710.04811 [hep-ph]).



\bibitem{Englert2020}
C.~Englert, D.J.~Miller and D.D.~Smaranda,
{\it ``Phenomenology of GUT-inspired gauge-Higgs unification''},
\jnl{\PLB}{802}{135261}{2020};
{\it ``The Weinberg angle and 5D RGE effects in a $SO(11)$ GUT theory''},
\jnl{\PLB}{807}{135548}{2020}.

\bibitem{Angelescu2022}
A.~Angelescu, A.~Bally, S.~Blasi and F.~Goertz,
{\it ``Minimal $SU(6)$ gauge-Higgs grand unification''},
\jnl{\PRD}{105}{035026}{2022}.

\bibitem{Maru2022}
N.~Maru, H.~Takahashi and Y.~Yatagai,
{\it ``Gauge coupling unification in simplified grand gauge-Higgs unification''},
\jnl{\PRD}{106}{055033}{2022}.

\bibitem{Angelescu2023}
A.~Angelescu, A.~Bally,  F.~Goertz and S.~Weber,
{\it ``$SU(6)$ gauge-Higgs grand unification: minimal viable models and flavor''},
\jnl{\JHEP}{04}{012}{2023}.  

\bibitem{MaruNago2024}
N.~Maru and R.~Nago,
{\it ``New models of $SU(6)$ grand gauge-Higgs unification''},
\jnl{\JHEP}{11}{035}{2024}.  


\bibitem{AnomalyFlow1}
S.~Funatsu, H.~Hatanaka,  Y.~Hosotani,  Y.~Orikasa and N.~Yamatsu,
{\it ``Anomaly flow by an Aharonov-Bohm phase''},
\jnl{\PTEP}{2022}{043B04 }{2022}, (arXiv:2202.01393 [hep-ph]).

\bibitem{AnomalyFlow2}
Y.~Hosotani,   
{\it ``Universality in anomaly flow''}, 
\jnl{\PTEP}{2022}{073B01 }{2022}, (arXiv:2205.00154 [hep-th]).





\bibitem{YHbook}
Y.~Hosotani,
{\it ``An Introduction to Gauge-Higgs Unification''},
ISBN 978-981-98-0097-1, World Scientific Publishing Company, Singapore (2025).
(doi.org/10.1142/14045)



\bibitem{Cacciapaglia2008}
G.~Cacciapaglia, C.~Csaki, J.~Galloway, G.~Marandella, J.~Terning and  A.~Weiler, 
{\it ``A GIM mechanism from extra dimensions''},
\jnl{\JHEP}{0804}{006}{2008}.

\bibitem{Adachi2010}
Y.~Adachi, N.~Kurahashi, C.S.~Lim and  N.~Maru,
{\it ``Flavor mixing in gauge-Higgs unification''},  
\jnl{\JHEP}{1011}{150}{2010}; 
Y.~Adachi, N.~Kurahashi and  N.~Maru,
{\it ``$D^0$-$\bar D^0$ mixing in gauge-Higgs unification''},  
\jnl{\JHEP}{1201}{047}{2012}; 
Y.~Adachi, N.~Kurahashi,  N.~Maru and K.~Tanabe,
{\it ``$B^0$-$\bar B^0$ mixing in gauge-Higgs unification''},  
\jnl{\PRD}{85}{096001}{2012}.


\bibitem{Falkowski2007}
A.~Falkowski,
{\it ``Holographic pseudo-Goldstone boson''},
\jnl{\PRD}{75}{025017}{2007}.

\bibitem{HS2007}
Y.\ Hosotani and Y.\ Sakamura,
{\it ``Anomalous Higgs couplings in the $SO(5) \times U(1)_{B-L}$ gauge-Higgs
unification in warped spacetime''},
\jnl{\PTP}{118}{935}{2007}.


\bibitem{PDG2024}
S.~Navas et al. (Particle Data Group),
{\it ``Reviews, Tables \& Plots''}, Chapter 12, 
\jnl{\PRD}{110}{030001}{2024}.



\bibitem{FCNCanalysis}
 A.\ Crivellin, G.\ D’Ambrosio, M.\ Hoferichter, and L.C.\ Tunstall, 
 {\it ``Violation of lepton flavor and lepton flavor universality in rare kaon decays}, 
 \jnl{\PRD}{93}{074038}{2016};
 F.\ Gabbiani, E.\ Gabrielli, A.\ Masiero, and L.\ Silvestrini, 
 {\it `` A complete analysis of FCNC and CP constraints in general SUSY extensions of the standard model''}, 
 \jnl{\NPB}{477}{321}{1996};
 P.T.P.\ Hutauruk, T.\ Nomura,  H.\ Okada and Y.\ Orikasa,
 {\it ``Dark matter and $B$-meson anomalies in a flavor dependent gauge symmetry''},
  \jnl{\PRD}{99}{055041}{2019}.



\bibitem{PDG2024b}
S.~Navas et al. (Particle Data Group),
{\it ``Summary Tables''}, Mesons, 
\jnl{\PRD}{110}{030001}{2024}.  



\end{thebibliography}
\end{document}